\documentclass[a4paper,fleqn,usenatbib]{mnras}

\usepackage{newtxtext,newtxmath}

\usepackage[T1]{fontenc}
\usepackage{ae,aecompl}

\usepackage{graphicx}	
\usepackage{amsmath}	
\usepackage{amssymb}	
\usepackage{pdflscape}

\title[An exceptional protostellar outburst]{Discovery of a mid-infrared protostellar outburst of exceptional amplitude}

\author[P. W Lucas et al.]{
P. W. Lucas$^{1}$, 
J. Elias$^{2}$, S. Points$^{3}$, Z. Guo$^{1}$, L.C. Smith$^{4}$, B. Stecklum$^{5}$, 
\newauthor  E. Vorobyov$^{6,7,8}$, C. Morris$^{1}$, J. Borissova$^{9,10}$,  R. Kurtev$^{9,10}$, C. Contreras Pe\~{n}a$^{11}$,   
\newauthor N. Medina$^{9,10}$, D. Minniti$^{12,13}$, V.D. Ivanov$^{14}$, R.K. Saito$^{15}$\\
$^{1}$Centre for Astrophysics, University of Hertfordshire, College Lane, Hatfield, AL10 9AB, UK\\
$^{2}$SOAR Telescope/NSFÕs NOIRLab, Casilla 603, La Serena, Chile\\
$^{3}$Cerro Tololo Inter-American Observatory/NSFÕs NOIRLab, Casilla 603, La Serena, Chile\\
$^{4}$Institute of Astronomy, University of Cambridge, Madingley Rd, Cambridge CB3 0HA, UK\\
$^{5}$Th\"{u}ringer Landessternwarte Tautenburg, Sternwarte 5, D-07778 Tautenburg, Germany\\
$^{6}$University of Vienna, Department of Astrophysics, Vienna 1180, Austria\\
$^{7}$Ural Federal University, 51 Lenin Str., 620051 Ekaterinburg, Russia\\
$^{8}$Research Institute of Physics, Southern Federal University, Rostov-on-Don, 344090, Russia\\
$^{9}$Instituto de F\'{i}sica y Astronom\'{i}a, Universidad de Valpara\'{i}so, ave. Gran Breta\~{n}a, 1111, Casilla 5030, Valpara\'{i}so, Chile\\
$^{10}$Millennium Institute of Astrophysics, Av. Vicuna Mackenna 4860, 782-0436, Macul, Santiago, Chile\\
$^{11}$School of Physics, University of Exeter, Stocker Road, Exeter, EX4 4QL, UK\\
$^{12}$Departamento de Ciencias F\'{i}sicas, Universidad Andr\'{e}s Bello, Fern\'{a}ndez Concha 700, Las Condes, Santiago, Chile\\
$^{13}$Vatican Observatory, V00120 Vatican City State, Italy\\
$^{14}$European Southern Observatory, Karl-Schwarszchild-Str 2, D-85748 Garching bei Muenchen, Germany\\
$^{15}$Departamento de F\'{i}sica, Universidade Federal de Santa Catarina, Trindade 88040-900, Florianopolis, SC, Brazil}

\date{Accepted XXX. Received YYY; in original form ZZZ}

\pubyear{2015}

\begin{document}
\label{firstpage}
\pagerange{\pageref{firstpage}--\pageref{lastpage}}
\maketitle

\begin{abstract}
We report the discovery of a mid-infrared outburst in a Young Stellar Object (YSO) with an
amplitude close to 8~mag at $\lambda \approx 4.6~\mu$m. WISEA~J142238.82-611553.7 is one of 23 highly
variable Wide-field Infrared Survey Explorer (WISE) sources discovered in a search of Infrared Dark Clouds 
(IRDCs). It lies within the small IRDC G313.671-0.309 ($d$$\approx$2.6~kpc), seen by the {\it Herschel}/Hi-Gal survey as a 
compact, massive cloud core that may have been measurably warmed by the event.
Pre-outburst data from {\it Spitzer} in 2004 suggest it is a class I YSO, 
a view supported by observation of weak 2.12~$\mu$m $H_2$ emission in an otherwise featureless red continuum spectrum 
in 2019 (6~mag below the peak in $K_s$). {\it Spitzer}, WISE and VISTA Variables in the Via Lactea (VVV)
data show that the outburst began by 2006 and has a duration $>13$~yr, with a fairly flat peak from 2010--2014.
The low pre-outburst luminosity implies a low mass progenitor. The outburst luminosity of a few $\times10^2$~L$_{\odot}$ 
is consistent with an accretion rate $\dot{M} \approx 10^{-4}$~M$_{\odot}$yr$^{-1}$, 
comparable to a classical FU Orionis event. The 4.6~$\mu$m peak in 2010 implies $T$ = 800-1000~K and a disc radial location 
$R\approx4.5$~au for the emitting region. The colour evolution suggests subsequent 
progression outward. The apparent absence of the hotter matter expected in thermal instability or MRI models may be 
due to complete obscuration of the innermost disc, e.g. by an edge-on disc view. Alternatively, disc 
fragmentation/infalling fragment models might more naturally explain a mid-infrared peak, though this is not yet clear.
\end{abstract}




\section{Introduction}

Eruptive variable YSOs are newborn star sytems wherein a sudden dramatic increase
in the accretion rate causes a large increase in flux, with a duration ranging from a few months up 
to a century \citep{herbig77, herbig89, hartmann96, audard14, cp17a}. The prototypical 
outburst was that of FU Ori \citep{herbig77} with a rise of 6 mag in the visible, thought to be 
associated with a thousand-fold increase in accretion rate for reasons not yet established (\citealt{hartmann96}). 
For a long time, this was the largest photometric outburst known but optical 
variations of 7~mag were seen in the atypical ``hot FUor" PTF14jg \citep{hillenbrand19}, identified
by the Palomar Transient Factory \citep{law09}. An initial 1 mag dip was followed by a 7 mag rise over 
150~days. Detection of such very high amplitude events may make it possible to rule out some of the 
physical mechanisms that have been proposed to explain them. There may however be more than one 
such mechanism, leading to events with different amplitudes and durations.

Traditionally, most high amplitude outbursts were detected in visible light but there
has been a recent trend towards detection of events in embedded YSOs in the near 
infrared \citep{cp17a}, in the submm/mm continuum
\citep{safron15, hunter17} and via methanol maser emission \citep{caratti17}.

In this paper we report the mid-infrared detection of a YSO outburst with a rise of 
$\sim$8~mag, higher than any previously known. This was a somewhat serendipitous discovery,
using data from the {\it WISE} satellite \citep{wright10}, with pre-outburst data from 
the Galactic Legacy Infrared Mid-Plane Survey Extraordinaire
\citep[GLIMPSE,][]{benjamin03}, by the {\it Spitzer} Space Telescope, \citep{werner04, gehrz07}.
It followed from our investigation of the unrelated infrared transient VVV-WIT-01 \citep{lucas20}, 
first detected by the VVV survey \citep{minniti10}. However, 22 additional high amplitude
mid-infrared variables were found in the same search (see section \ref{sec:method}).

While there have been past mid-infrared investigations of YSO variability \citep[see e.g.][]{rebull14, cp20} there 
have been only a few previous discoveries of outbursts in this waveband. 
\citet*{scholz13} found two likely eruptive YSOs (and a few less likely candidates) by comparing 
GLIMPSE and WISE data for known YSO candidates.  
\citet{antoniucci14} found a few candidate short duration outbursts of the EX Lupi type \citep[see][]{herbig77},
often referred to as EXors. by a similar comparison.
These previous mid-infrared discoveries all had amplitudes $\la$2~mag. By contrast, most of our discoveries
have mid-infrared amplitudes $\ga$2~mag. These higher amplitude findings demonstrate the 
potential of the ongoing {\it WISE}/NEOWISE mission for eruptive YSO science and the mid-infrared variable sky
more generally. Comparable high amplitude mid-infrared YSO outbursts have previously been detected only 
after discovery in another waveband. E.g. in the optically detected case of Gaia~17bpi \citep{hillenbrand18}, 
an optical and mid-infrared outburst was found to have been preceded by an earlier mid-infrared event with
no optical counterpart. \citet{stecklum20} recently discovered an outbursting YSO in WISE data by 
following up a report of a new optical nebula \citep{borisov20}.

The layout of this paper is as follows. In section 2 we describe our search for highly variable sources in the 
WISE data and briefly summarise the discoveries. In section 3 we present details of the extreme outbursting
YSO, including multi-wavelength public survey data and follow-up spectroscopy. We then
discuss what we can learn from the event in section 4, before giving our conclusions
in section 5.

\section{Variable star search}

\subsection{Method}
\label{sec:method}

We searched the WISE multi-epoch photometric catalogues for highly
variable sources and transients projected within two arcminutes of any of the 7139 IRDCs listed by 
\citet{peretto16} as bona fide dense molecular clouds. These IRDCs showed evidence 
for 250~$\mu$m emission in their automated analysis of images from the {\it Herschel}/Hi-Gal 
survey \citep{molinari10}. All the IRDCs are located in the {\it Spitzer}/GLIMPSE 
fields at Galactic lattitudes $|b|<1^{\circ}$ and longitudes $295^{\circ}<l<350^{\circ}$ or
$10^{\circ}<l<65^{\circ}$.

The search was undertaken during an investigation of a
very red transient, VVV-WIT-01, which was projected against an IRDC from the list of 
\citet{peretto09}, see \citet{lucas20}. The covering fraction of IRDCs in the GLIMPSE region
is only 1--2\% (depending on whether unconfirmed IRDCs are included) so 
it was thought that VVV-WIT-01 might be a pre-main sequence event. A search for additional 
transients in IRDCs therefore seemed logical, also offering the chance to detect
eruptive YSOs and other high amplitude variable stars. It would be preferable 
to search the available WISE datasets more fully (e.g. using the unWISE images, 
\citealt*{lang14, meisner18, schlafly19}) but this would be a computationally demanding task. The two arcminute 
search radius was chosen because it provides high completeness for any sources
inside IRDCs (98\% of the 7139 confirmed IRDCs have a smaller equivalent radius).
There are typically many WISE sources within a two arcminute radius in the
Galactic plane, even at the depth of individual WISE scans. 
The total area searched is $\sim$2~deg$^2$ which is only 1\% of the GLIMPSE area.
As part of its all-sky survey, the WISE satellite initially observed the GLIMPSE fields 
in 2010. The satellite was later reactivated as the NEOWISE mission and it observed the GLIMPSE fields 
from 2014 onward.

To construct a time domain catalogue in the WISE $W1$ (3.3~$\mu$m) and $W2$ (4.6~$\mu$m)
passbands, we downloaded from IPAC/IRSA all catalogue entries in the WISE All-Sky Single Exposure 
(L1b) Source Table located within the 7139 areas. We performed similar downloads from the
WISE 3-Band Cryo Single Exposure (L1b) Source Table, the WISE Post-Cryo Single Exposure (L1b) 
Source Table and the NEOWISE-R Single Exposure (L1b) Source Table. Together these comprise
data taken throughout 2010 and in the period from 2013 December to 2017 December. Since the 
WISE satellite scans the full sky every six months, each IRDC area typically has 10 epochs 
of observation (though more epochs have been added to the NEOWISE-R database since our
download). Each epoch is subdivided into several individual scans made over a $\sim$1~day interval,
for which photometry is recorded separately in the L1b source tables listed above.

We used the {\sc stilts} software \citep{taylor06} to perform initial matching of the catalogue entries, grouping them into 
time sequences for individual sources with a matching radius of 2\arcsec~(about a third of the WISE beam width). 
Duplicates of catalogue entries that were associated with multiple adjacent IRDCs were removed. 
Using a custom-written code, the per-scan magnitudes for detections in the $W1$ and $W2$ passbands and all the 
other associated catalogue parameters were then averaged to give values for each epoch. Then, starting with the 
epoch-averaged data for the 4-year NEOWISE-R dataset, the code cross-matched the sources associated with each 
IRDC against the epoch-averaged data for the same IRDC that were drawn from the three WISE datasets 
covering calendar year 2010. This provided a reasonably complete time domain catalogue for variable
stars. To improve completeness for transients, we repeated the 
cross-matching process using each of the three 2010 WISE datasets in turn to provide the initial source list.

In total, the resulting time domain catalogue contains a little over half a million sources. We investigated 
only the most promising high amplitude candidates. To search for variable stars, we selected
sources that varied by at least two magnitudes (maximum minus minimum) in at least
one of the two passbands ($W1$ and $W2$). (We later applied saturation corrections, see section \ref{sec:findings},
that reduced the amplitudes below 2~mag for a few of the brighter discoveries.) We then set quality criteria, to be satisfied 
in at least one passband, using parameters in the WISE L1b source tables. The criteria were designed to select unblended 
sources that were relatively bright and detected with a signal to noise ratio, $snr>5$ at one or more epochs.
An additional criterion related to the extent to which each source was well fitted by a point source model, 
based on relations between the fit quality parameter, $chi$, and signal to noise ratio, $snr$, as 
described in \citet{koenig14}. For this last criterion we averaged the $chi$ parameter in each passband
over all epochs with a detection and used the maximum value of the $snr$ parameter (after averaging these
parameters over the scans within each separate epoch) in order to not miss any bona fide detections.

The quality criteria can be written as: 
$\overline{NB} <1.1$ 
(i.e. the number of deblended sources contributing to the detection is $<1.1$, after averaging over all epochs) 
and the following conditions for the two passbands,

\begin{flalign*}
w1mpro<9~\mathrm{AND}~w1snr>5~~~~~~~~~~\mathrm{at~one~or~more~epochs}~~~~~~~~~~~~\\
\overline{w1chi} <5~\mathrm{OR} ~\overline{w1chi} < \frac{\left( \mathrm{max}\left(w1snr \right) -3\right)}{7}~~~~~~~~~~~~~~~~~~~~~~~~~~~~~~~~
\end{flalign*}
\vspace{-2mm}
\begin{flalign*}
w2mpro<9~\mathrm{AND}~w2snr>5~~~~~~~~~~\mathrm{at~one~or~more~epochs}~~~~~~~~~~~~~~~~~~~~~~~\\
\overline{w2chi}  <5 ~\mathrm{OR} ~ \overline{w2chi} < \left( 0.1~\mathrm{max}\left(w2snr \right) -0.3\right)~~~~~~~~~~~~~~~~~~~~~~~~~~~~~~~~~~~~
\end{flalign*}

\vspace{2mm}
\noindent Here $w1mpro$ and $w2mpro$ are the $W1$ and $W2$ magnitudes, $w1snr$ and $w2snr$ 
are the signal to noise ratios and $w1chi$ and $w2chi$ are the $\chi$ parameters, in the indicated
passband.

This selection yielded 177 candidates, all of which were visually inspected using the multi-epoch images
of the individual scans available at IPAC/IRSA. Of these, 14 were confirmed as bona fide high amplitude
variable stars. A further nine bona fide variables were found while testing different quality selections. These nine
failed the quality cuts on $\overline{NB}$ or the mean ({\it w}1/{\it w}2){\it chi} parameters due to poor data quality at one or 
two epochs. The 23 real high amplitude variable stars are listed in Table \ref{tab:table1}. This search is not 100\%
complete but merely a first exploration.

To search for transients we required similar quality criteria to be satisfied in at least one passband. We selected sources detected at only 
one epoch for which the number of deblended sources contributing to the detection, $NB$, satisfies $NB<1.5$. We applied 
essentially the same conditions on magnitude, $chi$ and $snr$ as above. (There was no need to average or maximise $NB$, $chi$ and $snr$ over 
multiple epochs, for these single epoch detections, though these quantities are averaged over the individual scans.)

This selection yielded 104 transient candidates, of which only one, VVV-WIT-01, passed
visual inspection. This transient search is likely to be more complete than our variable star
search, being less affected by data quality issues that affect only one or two epochs.

\subsection{Summary of variable star findings}
\label{sec:findings}

In Table \ref{tab:table1} we give the coordinates and mean $W1$ and $W2$ magnitudes of the 23 variable stars, together with the
amplitude in each passband (maximum minus minimum, after applying saturation corrections if needed\footnote{For saturation correction in 
$W1$ and $W2$ we used the tables at \url{http://wise2.ipac.caltech.edu/docs/release/neowise/expsup/sec2_1civa.html}}). We also give
an initial classification of each star as either a YSO or a candidate pulsating 
Asymptotic Giant Branch (AGB) star, based mainly on the WISE light curves and the 
VVV/VVVX\footnote{VVV \citep{minniti10, saito12} is a near infrared time domain survey of the Galactic bulge and the 
adjacent southern Galactic plane that ran from 2010--2015, using the 4-m Visible and Infrared Telescope for Astronomy \citep{sutherland15}. Time series 
data were taken in the $K_s$ bandpass. Multi-colour data was taken for each field in the $Z$, $Y$, $J$, $H$ and $K_s$ filters at two epochs, in 2010 and 
2015. VVVX is the continuing extension of the VVV survey, in duration and area \citep{minniti16}} light curves that are available for 20 of the 23. The candidate 
AGB stars are Long Period Variables, with periods of several hundred days. The very red colours (typically $1<W1-W2<4$) of the 23 variable stars imply 
that warm circumstellar matter is present in all cases, consistent with either a YSO or an AGB star with a high mass loss rate. While extinction in the IRDCs 
also reddens the colours, this cannot be the dominant effect because the mid-infrared extinction law is quite shallow in the WISE passbands 
\citep{koenig14} and the measured optical depth of the IRDCs has a mean of only 0.5 at 8~$\mu$m \citep{peretto09}. This 
optical depth corresponds to a colour excess of only $E(W1-W2)=0.17$, using the infrared extinction data from \citet{mcclure09} 
and \citet{koenig14}.

To search for periodic variability we included the additional time series data added recently to the NEOWISE-R database for 
calendar years 2018 and 2019, so that the full WISE time series covers 2010 and 2014-2019. A linear trend was then subtracted from the time series. 
(Slow trends are often present in mass losing AGB variables due to changing extinction by the circumstellar shell.) Then the {\sc IDL} implementation 
of the Lomb-Scargle normalized periodogram \citep{scargle82} was used to determine the frequency of the peak
frequency. Finally, a least-squares sine fit was used to refine it and provide amplitude and phase information (not shown). 
The peak to trough amplitudes of the sine fits to periodic sources range from 0.74 to 2.12~mag in $W1$ and similar values in $W2$.
We find that 13/23 stars are YSO candidates and 10/23 are AGB star candidates, ~the ~latter ~group defined by the detection of a long period and 
supported by their red colours. It is evident that the WISE mission provides a trove of valuable data for detection of highly variable 
\newpage

\begin{landscape}
\begin{table}
	\begin{centering}
	\caption{Visually confirmed high amplitude WISE variable stars in IRDCs}
	\label{tab:table1}
	\begin{tabular}{lccccccccccccccc} \hline	
No. & Name & RA$^a$  & Dec$^a$  & $\overline{W1}$ &  $\overline{W2}$ & $\Delta W1$ & $\Delta W2$ & Type & Period (d) & Notes and other identifications\\ \hline
  1  & WISEA J134444.02$-$623127.4 & 206.1835 & -62.5244 & 9.39  & 6.51  & 2.32 & 2.22 & AGB & 634.1$\pm$3.1 & [RMB2008] G309.0355-00.2858 \\
  2  & WISEA J142238.82$-$611553.7 & 215.6620 & -61.2650 & 8.05  & 5.80  & 1.87 & 1.54 & YSO & & Main subject of this work \\
  3  & WISE J142345.85$-$612540.7$^b$ & 215.9411 & -61.4282 & 13.01 & 11.44 & 3.74 & 3.67 & YSO & &    \\     
  4  & WISEA J154914.33$-$543423.6 & 237.3096 & -54.5733 & 11.41 & 9.53  & 2.86 & 2.18 & YSO & & 	\\	
  5  & WISEA J163957.05$-$462614.2 & 249.9877 & -46.4374 & 8.69  & 6.10  & 1.88 & 1.76 & AGB & 639.7$\pm$5.0	 & \\
  6  & WISEA J165035.47$-$444959.5 & 252.6479 & -44.8332 & 10.02 & 6.91  & 2.47 & 1.87 & AGB & 689.4$\pm$3.6 & [RMB2008] G340.7273-00.2234 \\
  7  & WISEA J165250.41$-$443908.4 & 253.2103 & -44.6524 & 10.59 & 7.02  & 1.83 & 1.85 & AGB & 760.9$\pm$6.4  & [RMB2008] G341.1209-00.4163 \\
  8  & WISEA J165344.39$-$432819.2 & 253.4351  & -43.4720 & 12.96 & 10.51 & 5.26 & 3.28 & YSO & & VVVv746, \\
  &  &  & &   &   &  &  &  &  &  [RMB2008] G342.1371+00.2054 \vspace{1mm}\\
  9  & WISEA J170547.35$-$411307.5 & 256.4473 & -41.2187 & 12.57 & 9.79  & 3.21 & 2.26 & YSO & & 	\\
  10 & WISEA J171910.90$-$390226.9 & 259.7952 & -39.0409 & 12.05 & 9.83  & 3.28 & 2.16 & YSO & & VVVv422 \\
  11 & WISEA J172258.05$-$370309.6 & 260.7421 & -37.0526 & 7.95  & 6.64  & 1.99 & 1.50 & AGB & 514.8$\pm$3.4 & IRAS 17195-3700\\
  12 & WISEA J181041.21$-$191040.2 & 272.6718 & -19.1778 & 11.09 & 9.34  & 2.97 & 2.31 & AGB & 371.6$\pm$4.8  & [RMB2008] G011.3064-00.0637 \\
  13 & WISEA J181426.60$-$172921.9 & 273.6110 & -17.4894 & 8.59  & 6.34  & 1.66 & 1.95 & YSO & & [RMB2008] G013.2154-00.0350, \\
  &  &  & &   &   &  &  &  &  &  YSO candidate in Marton et al.(2016). \vspace{1mm} \\
  14 & WISEA J181704.22$-$162554.0 & 274.2676 & -16.4318 & 9.72  & 6.69  & 2.01 & 1.80 & AGB & 891.0$\pm$7.5  &		\\
  15 & WISEA J181725.67$-$170211.7 & 274.3572 & -17.0366 & 12.24 & 8.61  & 3.25 & 2.09 & YSO & & [RMB2008] G013.9529-00.4460 \\
  16 & WISEA J181736.79$-$165006.2 & 274.4031 & -16.8351 & 8.37  & 7.00  & 2.50 & 1.73 & AGB & 501.7$\pm$7.3 &		\\
  17 & WISEA J181832.84$-$133239.3 & 274.6368  & -13.5443 & 11.54 & 9.23  & 2.61 & 2.60  & YSO & & [RMB2008] G017.1562+00.9715\\
  18 & WISEA J181849.10$-$140818.3 & 274.7048 & -14.1384 & 7.93  & 5.81  & 1.42 & 1.59 & AGB & 235.0$\pm$0.6  & [RMB2008] G016.6638+00.6324\\
  19 & WISEA J182025.44$-$163608.8 & 275.1059 & -16.6024 & 10.61 & 9.43  & 2.60  & 2.28 & YSO & & [RMB2008] G014.6746-00.8724\\
  20 & WISEA J182712.94$-$124904.8 & 276.8040 & -12.8180  & 7.57  & 5.89  & 1.33 & 1.62 & AGB & 788.8$\pm$8.3 & [RMB2008] G018.7877-00.5509\\
  21 & WISEA J185720.27+015711.8 & 284.3344 & 1.9534  & 7.88  & 6.96  & 3.47 & 3.23 & YSO & &  [RMB2008] G035.3429-00.4212,  \\
  &  &  & &   &   &  &  &  &  &  4.6 mag fainter in GLIMPSE. \vspace{1mm} \\
  22 & WISEA J190424.69+054106.8 & 286.1031 & 5.6853  & 10.03 & 8.59  & 2.03 & 2.01 & YSO & & Source 245 in Lucas et al.(2017),\\
  &  &  & &   &   &  &  &  &  & YSO candidate in Marton et al.(2016) \vspace{1mm} \\
  23 & WISEA J195146.18+272458.7 & 297.9423 & 27.4163  & 9.59  & 7.43  & 2.04 & 1.20 & YSO & & [RMB2008] G063.9380+00.2509,\\
  &  &  & &  &   &   &  &  &  &  YSO candidate in Marton et al.(2016).\\
  \\
 \multicolumn{16}{l}{Notes}\\
 \multicolumn{16}{l}{$a:$ The coordinates of source 21 given in decimal degrees are taken from UKIDSS, in order to distinguish this star from an optically brighter neighbour (see}\\
 \multicolumn{16}{l}{main text). All other sources have WISE-based coordinates from our time domain catalogue. All are equinox J2000 values.} \\
 \multicolumn{16}{l}{$b:$ Source 3 is not included in the AllWISE catalogue so the name is taken from the WISE All-Sky release.}
	\end{tabular}
	\end{centering}\\
\end{table}
\end{landscape}

\begin{figure*}
	\vspace{-4mm}
	\includegraphics[width=0.85\textwidth]{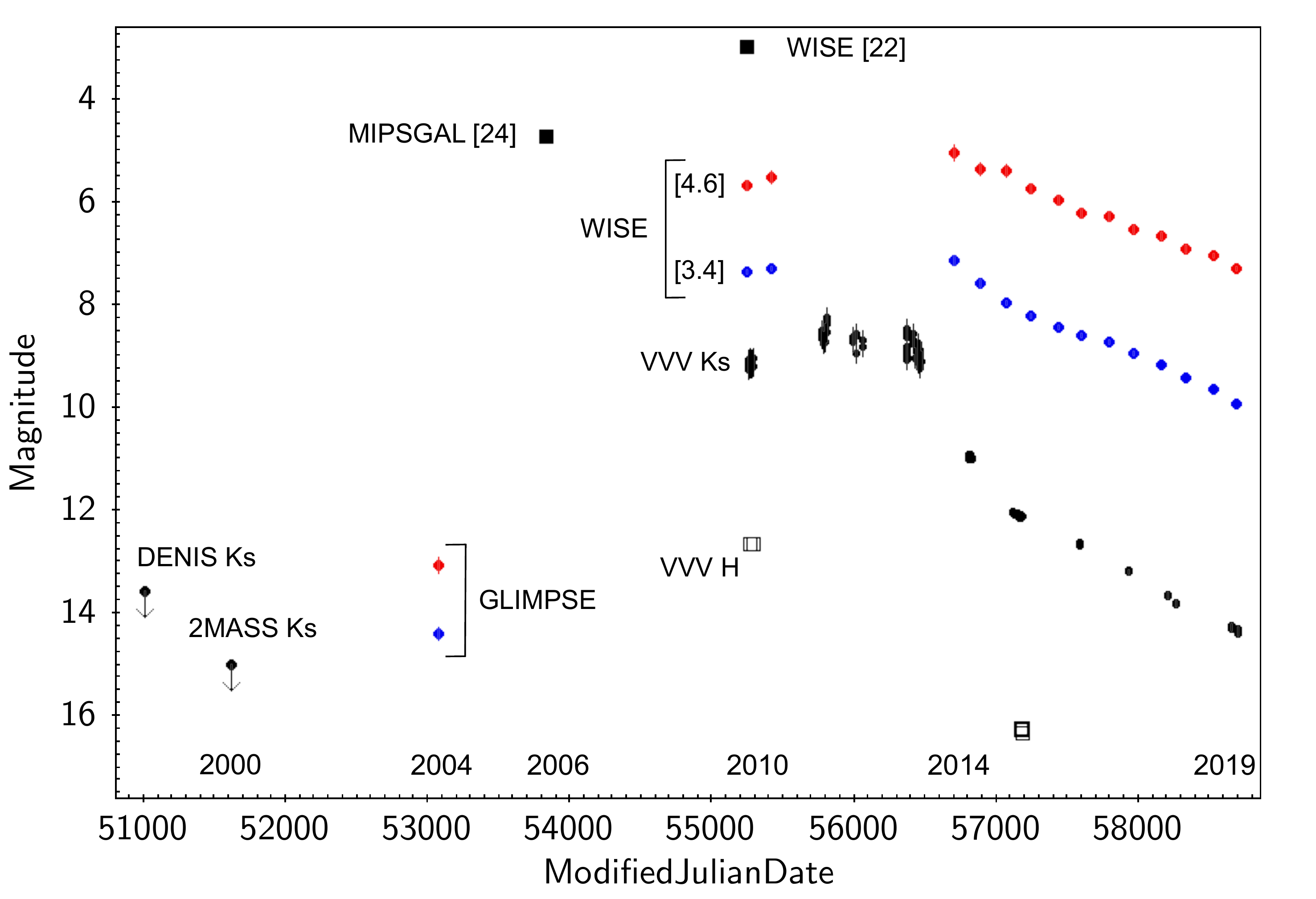}
    \caption{Multiwavelength light curves for WISE~1422-6115. The outburst appears to have begun between the GLIMPSE
    epoch in March 2004 and the MIPSGAL epoch in April 2006 (see main text). In 2019 the 3.4~$\mu$m, 4.6~$\mu$m and $K_s$ magnitudes remain 
    brighter than pre-outburst measurements and the 2MASS upper limit, implying that the outburst duration exceeds 13 yr. The 3.5 yr gap in the WISE data is due to 
    the inactive period of the satellite between the initial WISE mission in 2010 and the subsequent NEOWISE mission.}
    \label{fig:LC}
\end{figure*}

\noindent
YSOs in the Galactic plane. All the YSO candidates lie in the ``YSO region" of the $W1-W2$ vs. $W3-W4$ diagram that distinguishes YSOs from AGB stars, 
as defined separately by \citet{koenig14} and \citet{lucas17}, with the exception of source 2 in Table \ref{tab:table1}, see section
\ref{sec:discovery}. (We used the WISE All-Sky database to provide contemporaneous colours for this assessment).

YSOs dominate the Galactic population of high amplitude near infrared variable stars \citep{cp14,cp17a,lucas17}.
Given that our search was focussed on IRDCs we would therefore expect most of the 23 variable stars in 
Table~\ref{tab:table1} to be YSOs. However, this is a bright sample, which causes the proportion of AGB stars
to be relatively high. It is sometimes difficult to distinguish them from YSOs with periodic variability 
\citep[see][Guo et al., in prep.]{cp17a} but periodic YSOs are usually fainter.

Many of the sources were previously identified as YSO candidates in the infrared colour and 
SED-based searches of \citet[][sources designated ``RMB2008"]{robitaille08}, \citet{marton16} or by the time domain searches 
of \citep{cp17a, lucas17} using VVV and the UKIDSS Galactic Plane 
Survey \citep{lawrence07, lucas08}. Sources 13 and 23 in Table \ref{tab:table1} were listed as possible AGB stars by \citet{robitaille08} but classified 
as YSOs by \citet{marton16}. With the benefit of time domain data, we prefer the YSO classification for these aperiodic sources.

Source 2 in Table \ref{tab:table1} is the main subject of this paper due to the $\sim$8~mag difference between the
{\it Spitzer}/GLIMPSE and WISE magnitudes (see section \ref{sec:discovery}). However, one other star 
also showed a large change between the two surveys: source 21 brightened by 4.6 mag at 3.4~$\mu$m and 4.2~mag at 4.6~$\mu$m
between 2004 and 2014, after transforming the {\it Spitzer} photometry to the WISE passbands with the photometric transformations 
derived for YSOs in \citet{antoniucci14}. Most of this change occurred between 2010 and 2014 and the star has remained in a 
bright state since then. This object (SIMBAD identification [RMB2008] G035.3429-00.4212) was listed as a YSO candidate in \citet{robitaille08}.
In WISE, GLIMPSE and 2MASS \citep{skrutskie06} it is blended with an unrelated blue foreground star (which was listed in the 
catalogue of \citealt{vioque20} as a low-probability YSO candidate due to this source confusion). The two stars are clearly resolved
 in UKIDSS $K$-bandpass images from 2006, with a 0.9\arcsec~separation and similar brightness.
 They are not resolved in the second epoch UKIDSS $K$ images from 2011 because source 21 had become much brighter.

Separately, source 8, an eruptive YSO candidate listed previously by \citet{cp17a} as VVVv746, showed a $\sim$5.3~mag 
variation in $W1$ within the WISE data. This will be discussed in a forthcoming paper (Lucas et al., in prep., as a member
of the very high amplitude VVV YSO sample). The $\sim$5~mag infrared variations of sources 8 and 21 are towards the high 
end of what has been reported in classical FU Orionis events (FUors) and EXors (usually at optical wavelengths). 
We do not discuss these sources further here.

\begin{table}
	\begin{centering}
	\caption{Time series photometry for WISE~1422-6115}
	\label{tab:table2}
	\begin{tabular}{ccccccc} \hline
MJD & $W1$ & $\sigma_{W1}$ & $W2$ & $\sigma_{W2}$ & $W1-W2$ & $\sigma_{\mathrm{col}}$ \\ \hline
{\it 53075.5} &  {\it 14.41} & {\it 0.13} & {\it 13.08} & {\it 0.15} & {\it 1.33}  & {\it 0.20}\\
55243.6	& 7.38 & 0.04   & 5.70 & 0.08 &  1.86 &   0.09 \\
55422.2	& 7.30 & 0.04	& 5.52  & 0.13 & 1.78  & 0.14\\
56705.7	& 7.16 & 0.08     & 5.07 &  0.15  &  2.09   &  0.17 	\\
56886.9	& 7.58 & 0.07     & 5.37 &  0.14   &  2.21   & 0.16  \\
57070.4	& 7.98 & 0.03     & 5.42  &   0.14  &   2.56   &  0.14 \\
57245.7	& 8.23 & 0.03    & 5.75  &  0.11    &   2.48   &  0.11 	\\
57434.6	& 8.44 & 0.03   & 5.97  &  0.10     &   2.47   &   0.10 	   \\
57604.7	& 8.62 & 0.03     & 6.24 &  0.08    &   2.38   &  0.09  	\\
57798.8	& 8.73 & 0.03     & 6.29  & 0.08    &   2.44   &   0.09	\\
57967.3	& 8.97 & 0.03     & 6.55  &  0.07    &   2.42   & 0.08      	\\
58165.9	& 9.18 & 0.01      & 6.69  &  0.06     &  2.49    &  0.06    	\\
58328.5	& 9.44 &  0.01     & 6.93  & 0.05     &  2.51    &   0.05    	\\
58530.1	& 9.65 &  0.01     & 7.07 &  0.04     &  2.58    &   0.04    	\\
58692.7	& 9.94 &  0.01     & 7.31 &  0.04      &  2.63    &  0.04  	\\ \hline
\\
MJD & $W3$ & $\sigma_{W3}$ & $W4$ & $\sigma_{W4}$ & $W3-W4$ & $\sigma_{\mathrm{col}}$\\ \hline
55243.6	&   4.35    &  0.01 &   3.01  & 0.02 & 1.34 & 0.02	\\
55422.2	 &  4.24    & 0.03   &     & 	 & & 	\\ 
\\
\multicolumn{7}{l}{Note. Data in italics are from GLIMPSE, in $I1$ and $I2$. The}\\
\multicolumn{7}{l}{final column gives the uncertainty on the colour.}\\
\end{tabular}
	\end{centering}\\
\end{table}

\section{WISEA J142238.82-611553.7}
\subsection{Discovery and light curve}
\label{sec:discovery}

\begin{figure*}
	\includegraphics[width=\textwidth]{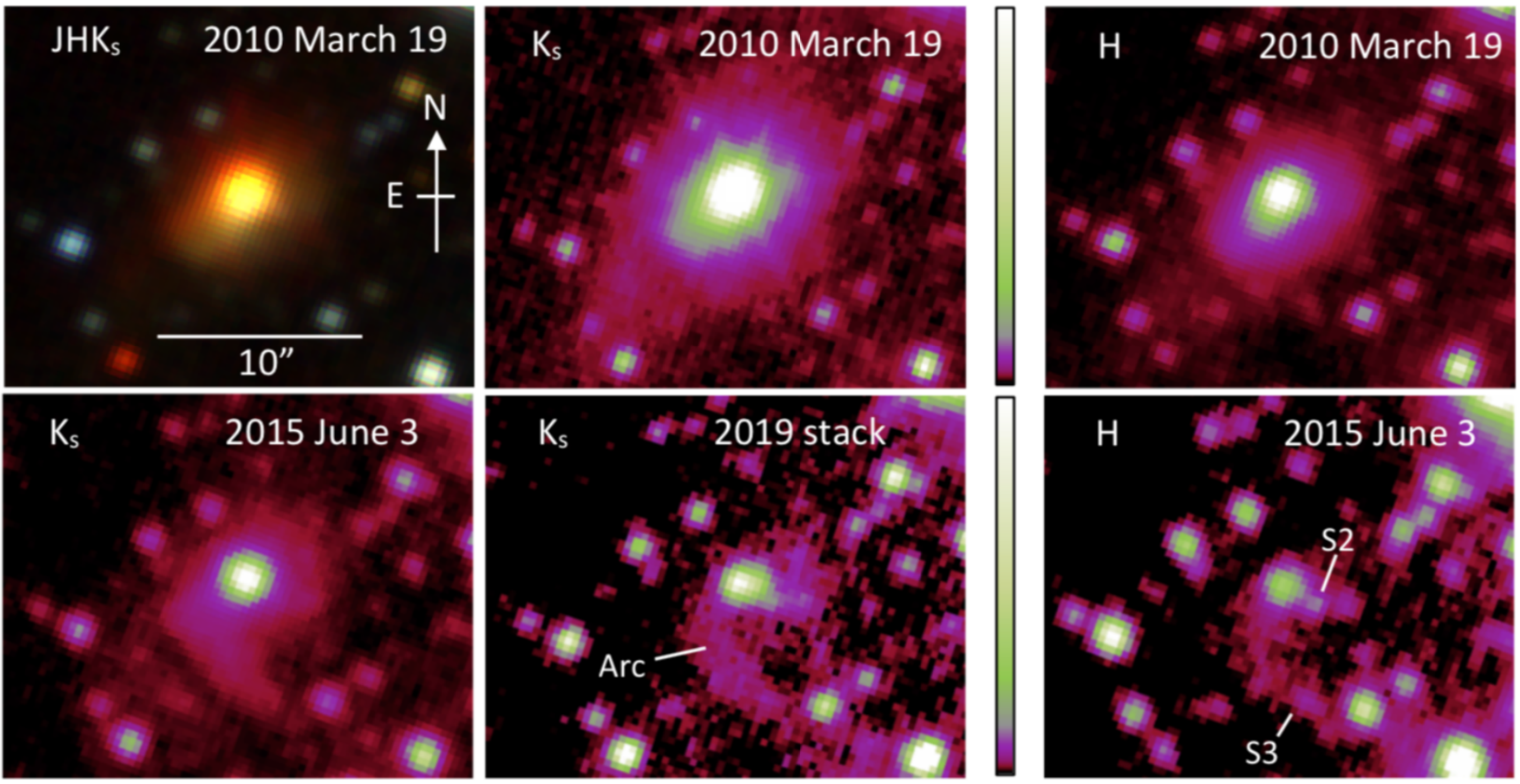}
    \caption{VVV near infrared Images WISE~1422-6115. All panels have the same spatial scale and equatorial orientation. The upper left panel is a three-colour image (red is $K_s$, green is $H$ and blue is $J$). This image taken in 2010, just below maximum brightness, shows the bright central flux peak and the faint cometary nebula extending to the southeast. The other five panels are false-colour images in the $K_s$ and $H$ passbands, scaled to show the changing morphology as the outburst faded (see main text). The $K_s$ and $H$ images use a colour map that emphasises faint features (see the vertical colour bars between the the centre and right panels). The images have a linear scaling from the sky background level to approximately the brightest count level in each panel.}
    \label{fig:VVVims}
\end{figure*}

Source 2 in Table \ref{tab:table1} was picked out as an object of special interest after cross-matching the 23 variable stars with the
{\it Spitzer}/GLIMPSE and GLIMPSE-II Point Source Archives \citep{churchwell09}.
Most sources have similar brightness (within 1~mag) in the $I2$ (4.5~$\mu$m) passband of the Infrared Array Camera (IRAC, \citealt{fazio04})
and the W2 (4.6~$\mu$m) passband but source 2 (=WISEA~J142238.82-611553.7, hereafter WISE~1422-6115) was brighter
by 8.0~mag at the brightest WISE epoch (2014 February 17) than it was at the GLIMPSE epoch (2004 March 11), see Table \ref{tab:table2}. At $\lambda \approx 3.5~\mu$m the rise in brightness from 2004 to 2014 was slightly smaller at 7.3~mag, based on a simple comparison
of the $I1$ (3.6~$\mu$m) and W1 (3.4~$\mu$m) magnitudes. If we transform the {\it Spitzer} $I1$ and $I2$ magnitudes to the WISE photometric system using the photometric transformations derived for YSOs in \citet{antoniucci14} then the $I2$ value becomes 0.20 mag brighter ($I2$=12.88) and the $I1$ value becomes 0.44 mag fainter ($I1=14.85$). The amplitudes at 3.4~$\mu$m and 4.6~$\mu$m are then more equal: approximately 7.7~mag and 7.8~mag respectively. 

Table \ref{tab:table2} lists the saturation-corrected\footnote{The saturation corrections (see section \ref{sec:findings}) made the brightest $W1$ and $W2$ data points become fainter by 0.28~mag and 0.47~mag, respectively. We give the error as the larger of the uncertainty on the saturation correction and the standard error on the mean derived from the individual scans at each epoch, the latter being an approximation to the standard error on the median. One highly discrepant scan was removed from the second $W3$ measurement.}
time series of multi-filter WISE photometry and 
the {\it Spitzer}/GLIMPSE $I1$ and $I2$ data, the latter given in italics using the original values from the GLIMPSE Point Source Archive because we do not know the uncertainty on the photometric transformation. For the WISE data in each passband we calculated the median magnitude of the several scans at each epoch, to minimise the effect of occasional outlying data points.

WISE~1422-6115 is projected within the small IRDC (SDC 313.671-0.309, equivalent radius 8.9\arcsec) from 
the list of \citep{peretto09}, at a distance 5.7\arcsec~from the centre. This IRDC has a compact far infrared and submillimetre counterpart 
in the form of a cold molecular cloud core, HIGALBM~313.6699-0.3092. The core was detected by the {\it Herschel}/Hi-Gal far infrared survey of the Galactic plane \citep{molinari10} and listed in the catalogue of \citet{elia17} as having a temperature of 15.8~K and a mass of 53~M$_{\odot}$. The presence of a large mass of cold matter centred within 4\arcsec~of the WISE source \citep[based on the Hi-Gal 250~$\mu$m coordinates given by][]{molinari16} helps to verify that this is a pre-main sequence event rather than an unlikely chance projection. There are 62,438 Hi-Gal sources in the ''high reliability`` catalogue of \citet{elia17}, widely spread across an area of 276~deg$^2$ at 
$-71^{\circ}<l<67^{\circ}$, $|b|<1^{\circ}$. The probability of a chance projection of an exceptional but unrelated event within 4\arcsec~of the listed position any of these sources is only $9\times10^{-4}$.
This conclusion is supported by other local star formation indicators (see section \ref{sec:envdist}) and the YSO-like morphology and spectrum (sections \ref{sec:nebula} and \ref{sec:spectrum}). As noted in section \ref{sec:findings} some of the WISE colours of WISE~1422-6115 (specifically the relatively blue $W3$-$W4$ value) resemble a mass-losing AGB star more than a YSO, according to the WISE two-colour diagram of \citet[][their figure 7]{koenig14}. However, it lies in the YSO region of the $W1$ vs. $W1$-$W2$ diagram in that work (their figure 9) and such colour-magnitude selections are known to be unreliable, as those authors  acknowledge. More importantly, the time domain behaviour (see Figure \ref{fig:LC}) does not resemble a pulsating AGB star. The longest period Mira variables have periods, $P < 6$~yr \citep{groenewegen20, navarro16}, typically with smooth, sinusoidal light curves. Unbiased searches of the near infrared variable sky with UKIDSS and VVV find that their peak to trough amplitudes very rarely exceed
$\Delta K_s = 4$~mag (\citealt{cp17a, lucas17}, Lucas et al., in prep.), whilst the mid-infrared amplitudes of the candidate AGB variables in Table~\ref{tab:table1} do not exceed 3~mag.

In Figure \ref{fig:LC} we plot the time series photometry and upper limits from {\it Spitzer}, WISE and the 2.15~$\mu$m data
from VVV/VVVX, 2MASS and the Deep Near Infrared Southern Sky Survey \citep[DENIS,][]{epchtein99}. These additional photometry are 
given in Table \ref{tab:A1}, along with mid-infrared photometry from the {\it Akari} all-sky survey \citep{ishihara10} and in Table \ref{tab:A2} the far 
infrared/submillimetre photometry from 
{\it Herschel}/Hi-Gal. The 2MASS image from 2000 March 12 provides an upper limit of $K_s=15.0$ (determined by inspecting the image and noting the 
faintest sources detected in the field in the 2MASS Point Source Catalogue and the 2MASS Survey Point Source Reject Table).

VVV/VVVX covers the 2010--2019 time period, consisting of 157 $K_s$ data points based on VVV pawprint image stacks.
WISE~1422-6115 is spatially resolved in VVV/VVVX images, showing a cometary morphology on a scale of several arc seconds, 
see Figure~\ref{fig:VVVims}. 
The source was also saturated near maximum brightness but fortunately the flux profile is strongly peaked so for comparison
with 2MASS we were able apply a simple saturation correction procedure to measure the brightness of the central flux peak, 
with $\sim$0.2~mag uncertainty, see Appendix \ref{sec:A}. The brightest VVV measurements were $K_s \approx 8.3$ on 2011 September 5. 
We also calculated the magnitude in an 8\arcsec~ diameter aperture on this date (see Appendix \ref{sec:A}) to give a best estimate of the integrated flux:
the integrated $K_s$ magnitude is only 0.05~mag brighter than the value quoted above. However, the cometary nebula makes a larger
contribution to the integrated $J$ and $H$ magnitudes (measured in 2010) which is relevant for SED modelling, see section \ref{sec:envdist}.

Inspection of Figure \ref{fig:LC} indicates that the outburst began between the GLIMPSE epoch in March 2004 and the epoch of the far infrared
{\it Spitzer}/MIPSGAL survey of the inner Galaxy in April 2006 \citep{rieke04, carey09}. The [8.0]-[24] colour is [8.0]-[24]=6.83 and the \citet{robitaille08} catalogue of intrinsically 
red {\it Spitzer} sources shows that this would be an extreme outlier for a non-variable YSO. Observation dates do not appear to be available for the 
{\it Akari} fluxes but the range of possible dates (May 2006 to August 2007) lies slightly after the MIPSGAL datum. The {\it Akari} data do not help us to determine the 
date of the outburst but the magnitude [9]=5.90 is 6.67 magnitudes brighter than the GLIMPSE 8~$\mu$m measurement, verifying that the outburst had a very high 
amplitude at $\lambda=8$--9~$\mu$m. In 2019 the WISE and $K_s$ magnitudes remain brighter than pre-outburst GLIMPSE measurements and the
2MASS upper limit, implying that the outburst duration exceeds 13 yr. The amplitude of the post-outburst decline is generally greater at shorter wavelengths.
The general trend towards higher amplitude at shorter wavelengths suggests that the amplitude 
in $K_s$ may therefore have been greater than the $\sim$8~mag change measured at $\lambda \approx 4.6~\mu$m. The 2MASS upper limit
tells us only that $\Delta K_s>6.7$~mag.  

The peak of the outburst was fairly flat from 2010--2013. However, fairly rapid $\sim$0.4~mag changes in the $K_s$ magnitude of the central peak
(confirmed by multiple measurements) did occur: over a one day interval in 2013 (MJD=56374--56375) and over a 23 day interval in 2011 leading up to the 
maximum observed brightness at MJD=55809.0. Variations at this level are common in embedded YSOs due to changes in extinction or accretion rate 
(if accretion is magnetically controlled, see \citealt*{romanova08}).

\subsection{Spatially resolved structure}
\label{sec:nebula}

Figure~\ref{fig:VVVims} shows that most of the extended nebulosity faded at approximately the same rate as the central flux peak, as expected for a reflection
nebula. Near maximum brightness (in 2010--2013) the cometary nebula extended $\sim$10\arcsec~to the southeast in $K_s$ and $H$. As the outburst faded in later years, the morphology changed. A faint arc of nebulosity curving to the south of the central source is seen in the 2015 $H$ image
and the 2019 $K_s$ image stack\footnote{Most of the single-band images in Figure~\ref{fig:VVVims} are cut-outs from VVV tile stack images, with 80~s time on source, taken during multi-colour observing blocks in 2010 and 2015 \citep{saito12}. Other VVV/VVVX observing blocks have shallower $K_s$ images so the ``2019" image was constructed from a stack of the 12 (8~s) $K_s$ pawprint stack images taken on three separate nights in 2019, giving a combined exposure time of 96~s.}, extending as far as the faint star labelled ``S3" in the 2015 $H$ image. This arc gradually becomes more visible
in the $K_s$ images from 2015 onwards, as the rest of the nebula fades. Other faint stars projected within the nebula
become visible as the outburst fades, e.g. the star labelled ``S2" adjacent to the central flux peak. (This is a fairly crowded Galactic star field and most stars in the images are probably in the foreground since they have relatively blue colours, see section \ref{sec:envdist}). The arc of nebulosity appears to have faded only slightly between 2015 and 2019 and it may have been present in 2010, simply hidden by the bright reflection nebula. Classical FUors are often associated with rings or arcs of optical or infrared reflection nebulosity \citep{goodrich87, liu16, takami18}. The latter two studies interpreted these features as evidence that YSO outbursts are triggered by gravitational instability, noting that long tail-like structures might be attributed to ejection of a massive clump of matter during a previous outburst. \citet{vorobyov20} modelled tail-like structures produced by ejected clumps or by dynamical encounters with passing stars, see also \citet{cuello20}, though the spatial scale of the modelled features was an order of magnitude smaller than that seen here.

If the arc of nebulosity lies at the rear of the cloud core, e.g. 0.15~pc behind the central source, then a light travel delay of 
$\sim$1~yr might explain why a dense filament of reflection nebulosity (for example) appears relatively bright when the rest of the nebula has faded.  Alternatively, this arc might possibly arise in shocked gas, in a situation where shocks created by interaction of high velocity and low velocity gas in an outflow produce UV line emission that is then reprocessed to lower frequencies. Its greater prominence in $H$ than in $K_s$ in the 2015 image suggests that it suffers less reddening than the rest of the nebula, tending to argue against the time delay interpretation.

The overall size of the circumstellar structure is surprising. The 2010 and 2015~$K_s$ images and the 2010 $H$ image show a cometary nebula of the kind commonly seen in nearby class I YSOs when light is scattered from the walls of a cavity in the circumstellar envelope, attributable to a molecular outflow \citep{whitney93}. However, the distance of the system is $d\approx2.6$~kpc (see section \ref{sec:envdist}) so the $\sim$10\arcsec~extent of the nebula to the southeast of the central source corresponds to 
26~\hspace{-1mm}000~au (0.13~pc). This is unrealistically large for a circumstellar envelope but not unusual for either a molecular cloud core or a molecular 
outflow. The fainter extremities of the nebula might arise from scattering by dust in an outflow \citep*{lucas97,chrysostomou07}. Alternatively, the outer portions of the Hi-Gal submm cloud core may be dense enough to be observable in scattered light.
The 2010 images and the 2015 $K_s$ image have a sudden drop in surface brightness 4 to 4.5\arcsec~southeast of the central peak. \citet{elia17} measured a beam-deconvolved radius of 4.5\arcsec~at 250~$\mu$m for the cloud core so the drop in $K_s$ surface brightness
at a similar radius might correspond a drop in the density of the cloud core. The alternative is that the drop in flux occurs at the edge of a rather large YSO envelope 
(10 000--12 000 au radius). The Hi-Gal radius measurement would appear to be uncertain due to the known issue of variation in beam-deconvolved source sizes as a function of wavelength \citep{elia17} and the fact that the size is much less than the {\it Herschel} beam size in this case. Nonetheless, the spatial scale resembles that of a cloud core better than a YSO envelope \citep[see][for an illustration of the hierarchy of size scales in star formation]{pokhrel18}.

\begin{figure*}
	\includegraphics[width=\textwidth]{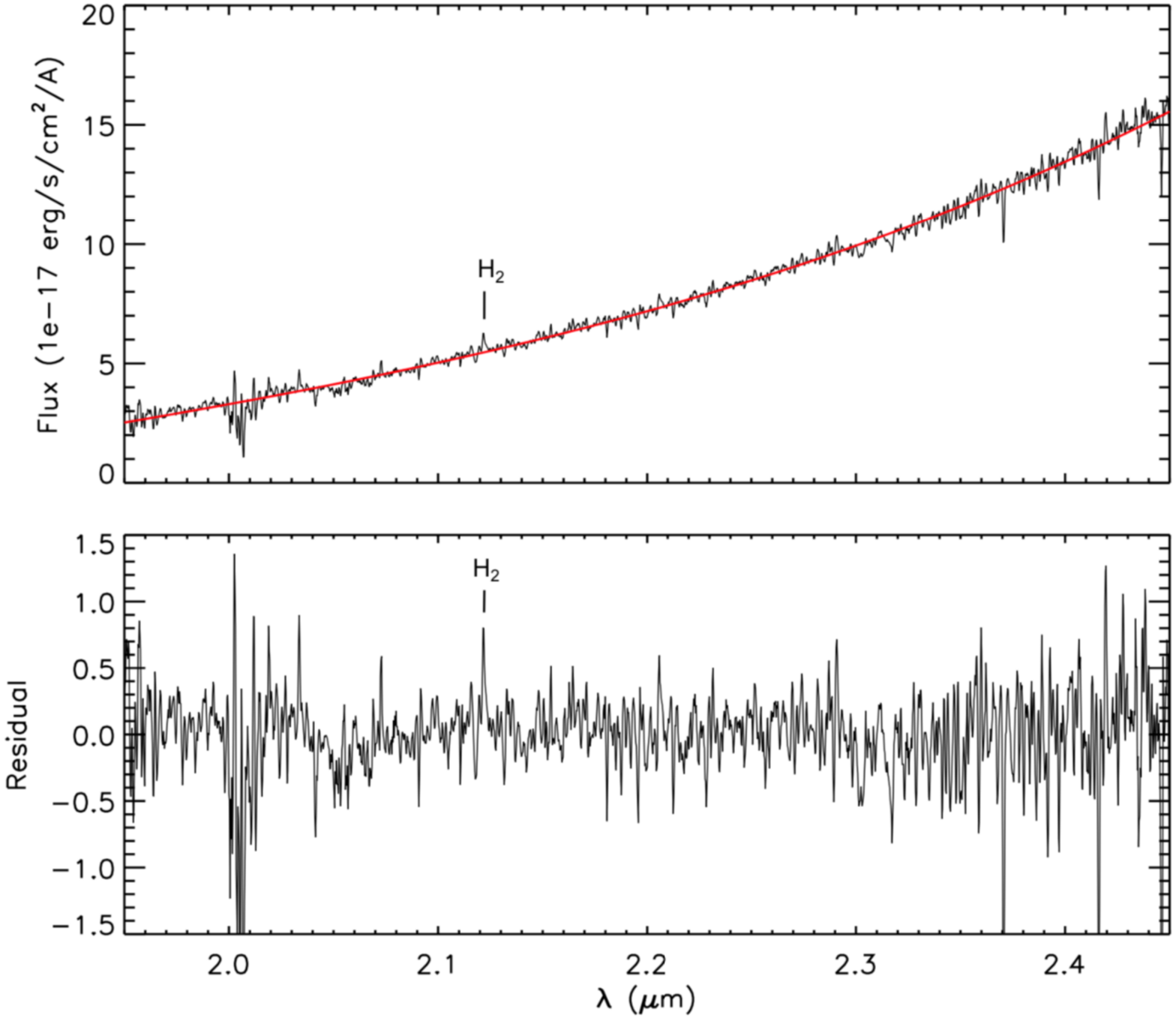}
    \caption{SOAR/ARCoIRIS spectrum of WISE~1422-6115. {\it (Upper panel)}: spectrum with the continuum fit by a cubic polynomial.  {\it (Lower panel)}:
    residual after subtracting the continuum fit. The (1-0) S(1) emission line of $H_2$ is weakly detected at 2.12~$\mu$m.}
    \label{fig:spec}
\end{figure*}

\subsection{Spectrum}
\label{sec:spectrum}

\begin{figure}
	\includegraphics[width=0.5\textwidth]{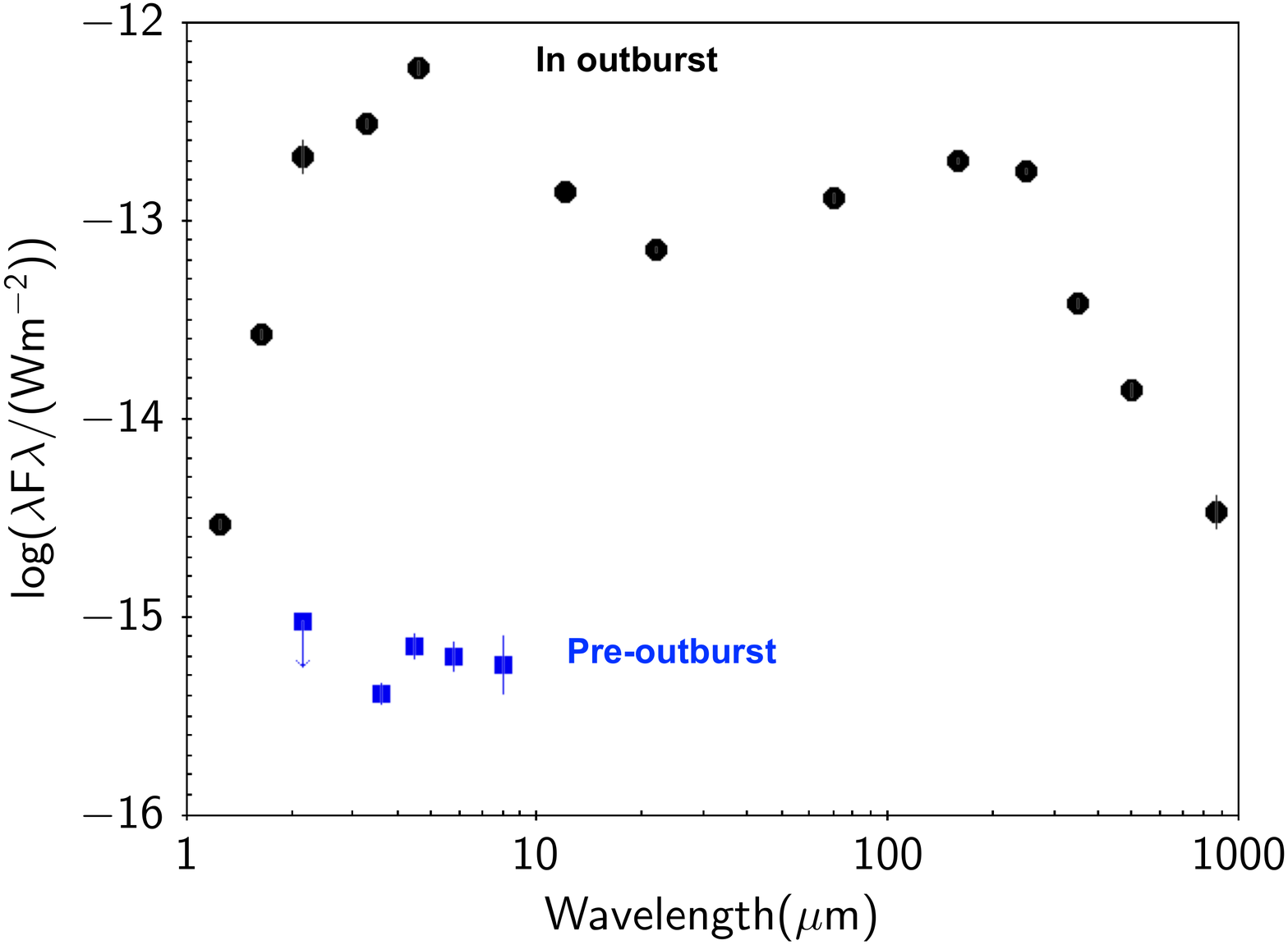}
    \caption{SED of WISE~1422-6115 and the surrounding cold cloud core (black points in outburst, blue points pre-outburst).
    The VVV, WISE and Hi-Gal in-outburst data are roughly contemporaneous, all taken in 2010, when the outburst was slightly below the peak. We also 
    include 870~$\mu$m ATLASGAL photometry from 2008. The form of this double peaked SED does not resemble the SED of a normal 
    embedded YSO: the near to mid-infrared peak is most likely due to a dramatic episodic accretion event in the inner disc whilst the far infrared/submm peak is 
    due to emission from a massive cold cloud core on a larger spatial scale. The latter peak is probably enhanced a little by the warming effect of the outburst 
    on the whole cloud core.}
    \label{fig:sed}
\end{figure}

An infrared spectrum of WISE 1422-6115 was obtained by co-author JE on the night of 2019 July 18/19 with the TripleSpec 4.1 infrared spectrograph on the 4-m Southern Astrophysical Research (SOAR) Telescope. TripleSpec 4.1 is a near-infrared cross-dispersed spectrograph \citep{schlawin14} providing a spectral resolution R=3500 across the 0.97-2.5~$\mu$m wavelength region, with a fixed slit width of 1.1 arcsec. Target acquisition with the instrument uses a slit-viewing system operating in the J-band where WISE 1422-6115 is too faint to detect. Consequently, the target was positioned on the slit by offsetting from the nearby field star. 

The integration time was 48 minutes on source, broken down into 16 three minute exposures. The spectra were reduced with the Spextool 4.1.0 pipeline, with independent reductions by co-authors SP and PWL providing essentially identical results. The A0V star HIP~76244 was used for telluric correction. The Spextool pipeline fits and removes the sky background in the dispersed images without the need for nodding the telescope between two positions and subtracting spectra. However, two separate nod positions were used in an ABBA pattern, enabling quick subtraction during the observations to confirm target acquisition and helping to guard against array defects.

The spectrum is shown in Figure \ref{fig:spec}, where we see a continuum that rises rapidly with increasing wavelength. There is a weak detection of
the (1-0) S(1) emission line of H$_2$ at wavelength $\lambda=2.1215~\mu$m, slightly blue-shifted from the rest wavelength of $\lambda=2.1218~\mu$m.
While the line is weak (the equivalent width is $-1.0 \pm 0.4$~\AA) it is detected in variously selected subsets of half of the data: position A observations, position B 
observations, the first eight exposures (i.e. the first two ABBA nod cycles) and the last eight exposures. Therefore it appears to be a genuine feature. 
There is no sign of the usual spectroscopic features associated with FUor or EXor outbursts (CO and H$_2$O in absorption in the former type, CO 
emission and Br$\gamma$ emission in the latter). This is not surprising because by 2019 the outburst had faded by 6~mag from its peak brightness in $K_s$.

The detection of weak 2.12~$\mu$m $H_2$ emission serves to support the interpretation of the system as a YSO because this line is typical
of class I YSOs \citep[e.g.][]{greene96,doppmann05} but rarely seen in other types of star. Other than YSOs, this emission line is
seen in protoplanetary nebulae or other sites of shocked gas such as supernova remnants but the former are not associated with high amplitude variability and this was clearly not a fast transient event. Amongst YSOs, the 2.12~$\mu$m line is much more commonly seen in class I YSOs than class II systems \citep{greene96}. The feature typically
arises in an outflow on a fairly large spatial scale so it tends not to correlate with accretion-driven photometric variations on timescales of only 1 or 2 years \citep{guo20}.

The radial velocity of the $H_2$ emission is estimated as $v_{\mathrm{LSR}}= -66 \pm 30$~km s$^{-1}$, after applying the heliocentric correction 
and correcting for the sun's motion relative to the Local Standard of Rest \citep*{schonrich10}. The uncertainty includes the line profile measurement error and the wavelength calibration uncertainty, with an allowance for an estimated 0.25\arcsec~uncertainty in the spatial location of the source within the 1.1\arcsec~slit.
This radial velocity is broadly consistent the system velocity of $v_{\mathrm{LSR}} \approx -40$~km s$^{-1}$ that we expect, based on our
our preferred distance, $d \approx 2.6$~kpc see section \ref{sec:envdist}). However, we note that $H_2$ emission from a YSO outflow can have a velocity that is 
significantly offset from the system's reference frame \citep[e.g.][]{guo20}.

\subsection{SED}
\label{sec:SED}

In Figure \ref{fig:sed}, we show the SED of WISE~1422-6115 and the surrounding cold cloud core from 1~$\mu$m to 1~mm. Blue points are the data from GLIMPSE 
and the 2MASS upper limit, all taken pre-outburst. Black points are based on data taken during the bright part of outburst: VVV, WISE and {\it Herschel}/Hi-Gal data were all taken in 
2010, when VVV $K_s$ and WISE $W2$ fluxes were only a factor of $\sim$2 lower than the highest values measured. The VVV and WISE data points were taken only a month apart 
(February to March 2010) but the Hi-Gal data from \citet{elia17}, for the core designated HIGALBM~313.6699-0.3092, are an average of several observations at more separated dates in 
the year. The 870~$\mu$m datum was taken in 2008, also during the bright stage of the outburst (see Figure~\ref{fig:LC})
by the 12-m Atacama Pathfinder Experiment (APEX) Telescope \citep{gusten06}
as part of the APEX Telescope Large Area Survey of the Galaxy (ATLASGAL, \citealt{schuller09}). The Hi-Gal and ATLASGAL fluxes at $\lambda>250~\mu$m were scaled to the 
measured size of the 250~$\mu$m detection, a procedure adopted by \citet{elia17} for SED 
fitting for all the molecular clumps and cores in their catalogue (see that work and references therein). The VVV $J$, $H$ and $K_s$ photometry shown here are integrated fluxes 
measured in an 8\arcsec~diameter aperture, for better comparison with the lower resolution WISE photometry.\footnote{The integrated VVV magnitudes ($J$=15.31, $H$=12.18, 
$K_s$=9.11, based on data from 2010 March 19) are significantly brighter in $J$ and $H$ than the point source magnitudes given in Table~\ref{tab:A1} ($J$=16.06, $H$=12.65, 
$K_s$=9.17) because the cometary nebula makes a substantial contribution up to a 4\arcsec~radius, though not beyond that.}

\subsubsection{Pre-outburst SED}
The pre-outburst GLIMPSE colours suggest that WISE~1422-6115 is a class I YSO: $I1$-$I2$=1.33, $I3$-$I4$=0.87 and $I2$-$I4$=1.51 are typical of class I YSOs in 
\citet{gutermuth09}. The $I2$-$I3$=0.64$\pm$0.23 measurement is blue for a class I system, but the uncertainty allows it to be consistent with the colour space 
adopted by those authors. These colours add to the evidence for class I status provided by the 2.12~$\mu$m H$_2$ emission line mentioned earlier. Moreover, the
location within a dense cloud core provides circumstantial evidence that this is a protostar rather than a more evolved pre-main sequence star.

\subsubsection{In-outburst 1--22~$\mu$m SED}
\label{sec:burst_sed}

In outburst, the 1--22~$\mu$m SED can be fitted by a range of stage 2 (envelope-free) YSO models with the \citet{robitaille07} model grid (Python {\sc sedfitter} version) but the far 
infrared signature of any envelope that may be present is obscured by the cloud core.
The decline in the (in-outburst) flux from 4.6~$\mu$m to 22~$\mu$m indicates that the outburst takes place in the inner disc and suggests that radiation is not very efficiently transferred 
to larger disc radii. 
Alternatively, there might be a gap in the disc. Using $\lambda T=3.67$ mm K at the peak of a Planckian SED, the infrared
peak at 4.6~$\mu$m implies $T\sim800$~K. The in-outburst infrared luminosity measured in 2010 from 1.25~$\mu$m to 22~$\mu$m was $L=178 (d/2.6 \mathrm{~kpc})^2$~L$_{\odot}$. 
We calculated this very simply by the Trapezium Rule, i.e. treating the flux in the wavelength interval between each pair of points in the SED as the average of the two flux densities, multiplied by the wavelength difference. Using these values for $L$ and $T$, the equation $L=2 \pi R^2 \sigma T^4$ yields a radius $R~=~4.6$~au for the outburst. 

We might expect there to be significant infrared extinction towards the inner disc of a class I YSO that is located inside an IRDC/dense cloud core.
An experiment with extinguished single-temperature black body fits to the 1.25--22~$\mu$m SED (using $\chi^2$ fitting) found that an 800~K source with zero extinction is the best
fit (with 40--60\% residuals at some wavelengths) but models with extinction up to $A(K_s)=1.5$~mag and temperatures up to 1000~K are only slightly worse, raising the 
$\chi^2$ parameter by only 60\%. (We used the near infrared extinction law of \citealt{cardelli89} for the VVV data and the mid-infrared extinction law 
given by \citealt{koenig14} for WISE data in fields with high extinction.) Adopting $T$=800--1000~K and $0<A(K_s)<1.5$ implies $L$=178--418 $(d/2.6 \mathrm{~kpc})^2$~L$_{\odot}$, 
from 1.25--22~$\mu$m, and the radius of the emitting region 
remains tightly constrained at $R$=4.5--4.6~au. The fits to the \citet{robitaille07} model grid provide some guidance here, even though these models were not designed to accommodate 
YSO outbursts. In the best-fit models, the foreground extinction is $15<A(V)<26$~mag and the disc is observed edge-on (system inclination $i=87^{\circ}$) such that the star is entirely 
hidden from view by the disc and only the disc contributes directly to the observed SED. (In these models, the stellar photosphere has a temperature $5700<T<7600$~K, which happens 
to be similar of the emitting region of the disc measured in optical spectra of classical FUors.)

\subsubsection{The Hi-Gal cloud core}

\begin{figure}
	\hspace{-6mm}
	\includegraphics[width=0.55\textwidth]{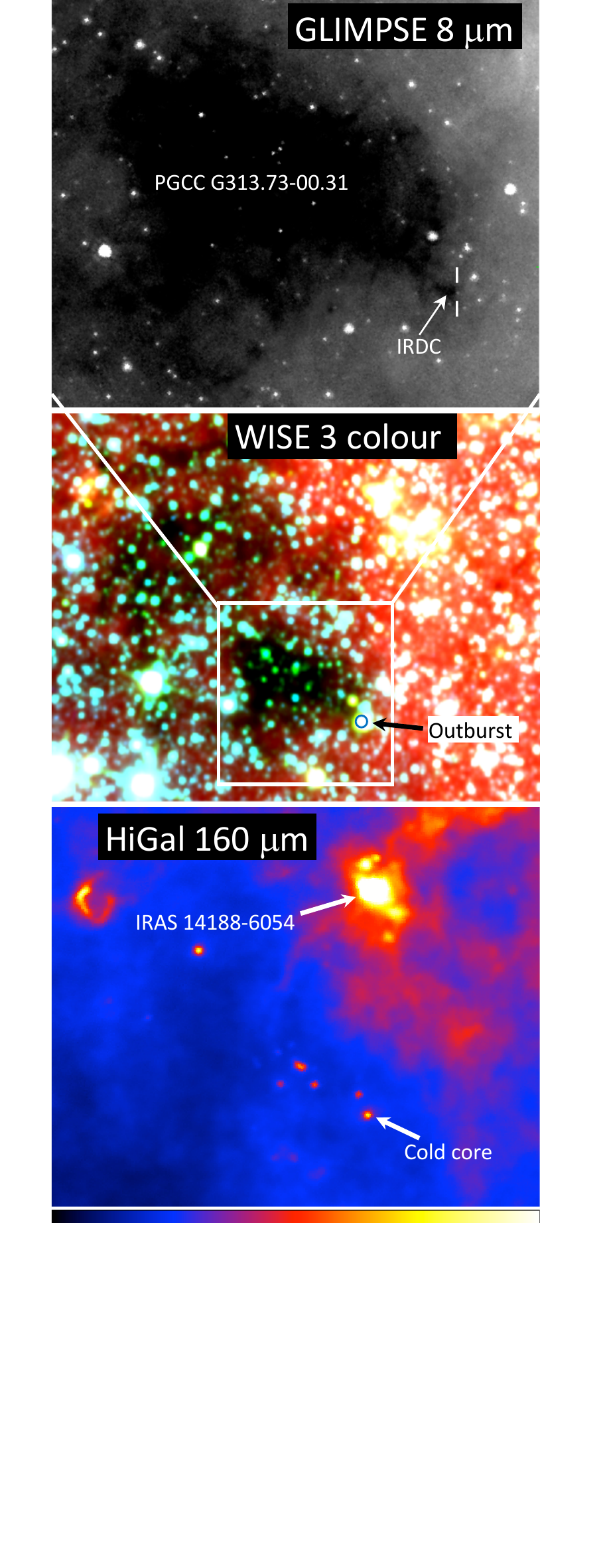}
    \vspace{-5.8cm}
    \caption{The wider environment. All images have equatorial orientation. {\it (top:)} GLIMPSE 8~$\mu$m 6$\times$5 arcminute image taken pre-outburst, 
    in 2004. WISE~1422-6115 is faintly seen, between the vertical white markers. The IRDC (G313.671-0.309) is the adjacent small dark patch and the Planck cloud 
    PGCC G313.73-00.31 dominates the image as a dark region. {\it (middle:)} WISE $16 \times 12$ arcminute three colour image (blue: 3.4~$\mu$m, 
    green: 4.6~$\mu$m, red: 12~$\mu$m) taken in 2010. The erupting YSO (marked with an arrow and a blue circle) was prominent. {\it (bottom:)} Hi-Gal 
    160~$\mu$m $16 \times 12$ arcminute false colour image, with a linear stretch (see colour bar below). The IRAS 14188-6054 HII region and the cold cloud core 
    surrounding WISE~1422-6115 are marked.}
    \label{fig:env}
\end{figure}

The Hi-Gal cloud core appears very compact in all the Hi-Gal 70 to 500~$\mu$m images (see Figure \ref{fig:env}). \citet{elia17} fit a beam-deconvolved diameter of 
9.09\arcsec~at 250~$\mu$m. Their grey-body SED fit to the 170 to 870~$\mu$m Hi-Gal and ATLASGAL data gives a mass $M=53~M_{\odot}$, temperature $T=15.8 \pm 0.2$~K and 
optical depth unity at wavelength $\lambda_0=81.5~\mu$m. (Their adopted distance, 3.13~kpc, is comparable to our preferred distance of 2.6~kpc). This high mass and cold 
temperature make it clear that at wavelengths $\lambda>100~\mu$m the SED is dominated by a cloud core rather than a circumstellar envelope. The double-peaked SED, 
with the higher peak in the near to mid-infrared, does not resemble a normal class I or class 0 YSO: we confirmed this by trying unsuccessfully to fit it with the \citep{robitaille07} SED 
fitting tool. However, as discussed below (section \ref{sec:reproc}), heating by the outburst may well have increased the Hi-Gal and ATLASGAL cloud core fluxes somewhat.

We regard the beam-deconvolved diameter of the core as uncertain since it is much smaller than the {\it Herschel}/Hi-Gal beam at 250~$\mu$m (see \ref{sec:nebula}).
 (The published value is consistent with the size of the darkest portion of the IRDC as it appears in the GLIMPSE $I4$ image but smaller by a factor of two than the IRDC equivalent radius given by \citealt{peretto16}). Therefore we performed our own grey-body fits of the 170 to 870~$\mu$m SED to check the published parameters, in a similar manner to \citet{ward-thompson90}. We initially added the core's solid angle on the sky, $\Omega$, and then also the opacity power law index $\beta$ as free parameters \citep[$\beta=2$ was adopted by][]{elia17}. We found that $\beta$, $\Omega$ and $\lambda_0$ were not well constrained by the data but the temperature is well constrained, with minimum $\chi^2$ values corresponding to $T=15$~K for a 4 parameter fit, or $T=16$~K if we set $\beta=1.8$. These temperatures are in close agreement with \citet{elia17}. Temperature is the key parameter that determines the derived mass (when distance is known) so our mass estimates were of the same order as that of \citet{elia17}.

The mid-infrared decline in the SED is unusual for Hi-Gal protostellar cores with AllWISE counterparts in the \citet{elia17} catalogue. Using their definition of cores as bodies with radii 
$r <0.1$~pc, we found that during outburst WISE~1422-6115 was a blue outlier with $W2-W4=2.68$, compared to $4<W2-W4<9$ for most other 
cores. Unsurprisingly, it is one of the brightest sources in $W2$ (4.6~$\mu$m).\footnote{Our cross-match to AllWISE found eight Hi-Gal ``protostellar cores" with much brighter and somewhat bluer WISE counterparts than WISE~1422-6115 ($W2<0$, $W2-W4<2.5$). However, seven of these are listed in the SIMBAD database as evolved stars, e.g. OH/IR stars. Only one, CRL~2136, is a protostellar object.}

\subsubsection{Reprocessing}
\label{sec:reproc}

The higher luminosity of the shorter wavelength peak is consistent with YSO outburst models \citep{johnstone13, macfarlane19}, wherein an outburst in the inner disc greatly raises the near and mid-infrared flux but has a gradually decreasing effect at longer wavelengths. In their model, radiation is reprocessed in the circumstellar envelope (assuming there is one) with 
a warming effect. However, the Rayleigh-Jeans tail of a Planckian SED flux scales only linearly with temperature, leading to a smaller
effect at submm wavelengths. In WISE~1422-6115, the pre-outburst infrared luminosity (blue points in Figure~\ref{fig:sed}) is negligible compared to the luminosity of 
the massive cold cloud core. The in-outburst infrared luminosity ($L$=178--418~$(d/2.6 \mathrm{~kpc})^2$~L$_{\odot}$ from 1.25--22~$\mu$m measured in 2010) exceeds the cold core's luminosity from 70--870~$\mu$m ($L=59 (d/2.6 \mathrm{~kpc})^2$~L$_{\odot}$) by a factor of three to seven, perhaps even more at the $K_s$ peak in 2011. The core is therefore likely to have been heated by a few kelvins but this would not be expected to cause significant chemical changes.
 The heating efficiency would depend somewhat on whether the radiation is first reprocessed by a smaller circumstellar envelope and re-emitted at submm wavelengths where the optical depth of the core is low; future measurements of the core's SED would should reveal the size of the effect. 

The 2010 luminosity measured from the 1.25~$\mu$m to 870~$\mu$m SED is $L$=263--503$~(d/2.6 \mathrm{~kpc})^2$~L$_{\odot}$, adopting $0<A(K_s)<1.5$. The ratio of pre-outburst to in-outburst luminosity in the 3--5~$\mu$m region is $\sim$850 (comparing the WISE $W1$ and $W2$ data in March 2010 to the GLIMPSE magnitudes given in section \ref{sec:discovery}, after transforming to the WISE photometric system). If this flux ratio is applied to the whole SED, the implied quiescent luminosity is only 0.3--0.6~L$_{\odot}$. The ratio of far infrared to mid-infrared luminosity during quiescence may have been an order of magnitude higher than in 2010, assuming a class I SED, giving a quiescent luminosity of a few L$_{\odot}$. Even so, it is clear that the progenitor is a low mass YSO.

\begin{figure*}
	\includegraphics[width=\textwidth]{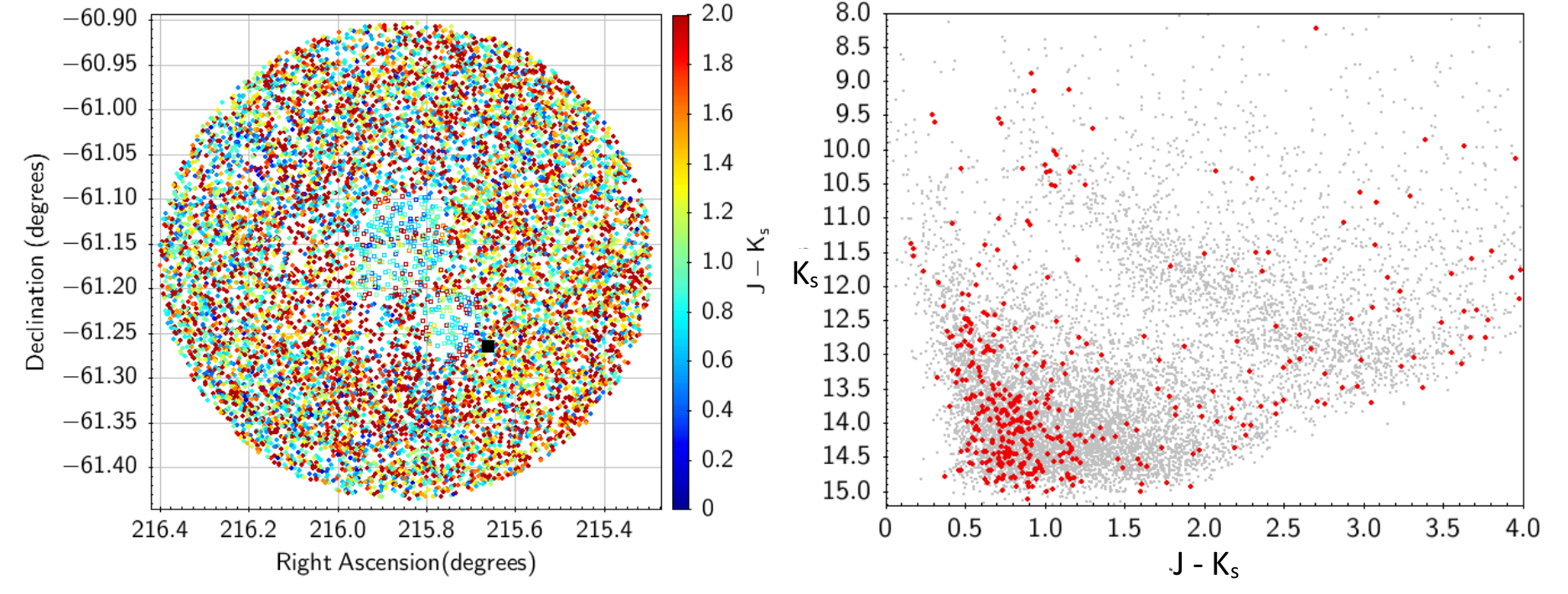}
    \caption{Red clump giant extinction-based distance to the Planck cloud PGCC G313.73-00.31. {\it (left:)} 2MASS map of sources in a 16 arcminute radius area centred on
    the region of highest extinction. The colour coding according to $J-K_s$ colour shows that very few red stars are visible near the centre, where the cloud is opaque. Stars 
    are plotted with open squares in the selected opaque (blue-ish) region, where foreground stars predominate. The location of WISE~1422-6115 is marked with a black 
    square.  {\it (right:)} 2MASS $K_s$ vs. $J-K_s$ CMD for the area of the left panel. Stars in the selected opaque (blue-ish) region are plotted in red and the rest in grey,  
    providing a large control area. The giant branch is well defined by the grey points but the distribution of red points in the giant branch cuts off abruptly near $K_s$=10.5. 
    We infer a red clump giant distance $d \approx 2.6$~kpc to the dark cloud and apply this distance to WISE~1422-6115 also.} 
    \label{fig:cmd}
\end{figure*}

\subsection{Wider environment and distance}
\label{sec:envdist}

In Figure \ref{fig:env} we illustrate the star forming environment around WISE~1422-6115. In the upper panel (a GLIMPSE 8~$\mu$m image taken pre-outburst),
WISE~1422-6115 is seen as a very faint source at the edge of the IRDC G313.671-0.309, which in turn appears as an unremarkable small patch at the edge of a much larger dark cloud. We identify this large cloud as PGCC G313.73-00.31 from the list of ``Planck Galactic Cold Clumps" \citep{ade16}. Those authors calculate distances to the molecular clouds in their list using several methods, finding $d=2.72 \pm 0.82$~kpc for PGCC G313.73-00.31. That was an extinction-based distance derived from a comparison of the colours and magnitudes of all stars in the region with the 
prediction of the Besan\c{c}on Galactic stellar population models \citep[][see \url{https://model.obs-besancon.fr}]{robin03}. They noted that the result was consistent with a near kinematic distance of 3.15 kpc for the IRDCs in the vicinity \citep{jackson08}: in fact the latter authors find distances ranging from 2.76 to 3.15~kpc for different components of the G313.70-00.31 IRDC, with
CS (2-1) velocities $-44.3 < v_{LSR} < -39.4$~km~s$^{-1}$. 
Other authors have measured similar radial velocities for cold cloud cores within the Planck clump \citep{vasyunina11, purcell12},
and the former work specifies a distance $d\approx 3.3$~kpc. 

The middle and lower panels of Figure \ref{fig:env} give a wider field view (16 arcminute width). The dark cloud is seen to extend further to the northeast in WISE three colour image (the whole dark area corresponding to dark cloud no. 6082 of \citealt{dobashi11}). WISE~1422-6115 appears bright in  the WISE image and the Hi-Gal cold cloud core (HIGALBM~313.6699-0.3092) is seen at the same location in the lower panel.

Other bright regions in the middle and lower panels of Figure \ref{fig:env} correspond to star formation activity at a kinematic distance similar to those mentioned above.
E.g. IRAS~14188-6054 is an HII region containing a 6.7~GHz methanol maser, MMB G313.705-00.190 \citep{green12}, and the associated ATLASGAL cloud core  
AGAL~313.706-00.191. The ATLASGAL team preferred a near kinematic distance of $3.0^{+0.6}_{-0.5}$~kpc \citep{wienen15} whereas \citet{whitaker17} preferred the far kinematic distance of 8.5~kpc. 

Figure \ref{fig:cmd} illustrates our own extinction-based distance calculation, a red clump giant distance \citep{lopez-corredoira02} based on the giant branch as measured in the $K_s$ vs $J-K_s$ colour magnitude diagram (CMD) constructed from the 2MASS Point Source Catalogue. (VVV data could not be used for this because the
cloud is too nearby, causing foreground stars at the extinction discontinuity to saturate). We plotted a map of three-filter detections ($J$, $H$, $K_s$), shown in the
left panel of Figure \ref{fig:cmd}, and found that the dark cloud area seen in the middle panel of Figure \ref{fig:env} corresponds to an opaque region where very few red stars are present. This region appears blue in the detection map due to the predominance of foreground stars with relatively blue colours. We traced this ``blue region"
by hand due to its irregular shape (in fact it is split into two adjacent areas separated by a narrow, slightly less opaque region populated by red stars). Stars in the 
selected blue region are plotted with open squares in the detection map. In the right panel of Figure \ref{fig:cmd}, stars in the selected region are plotted as red circles.
We see that the giant branch is very sparsely populated, due to a sudden cut-off near $K_s=10.5$~mag. By contrast, in the larger control region enclosing the selected 
area (16 arcminute radius centred on (ra, dec)=(215.8471, -61.1694), data plotted as small grey circles) the giant branch is well defined and continuous. We use
this larger area as a control field to estimate the average population of red clump giants per unit magnitude in the giant branch because smaller areas have differing levels 
of extinction, leading to scatter in the results. 

After scaling the area of the control field to the selected high extinction area of the dark cloud, we adopt the red clump giant colour, absolute magnitude and extinction law employed by \citet{minniti11} and we find that the cut-off occurs between 2.3 and 3.0~kpc (i.e. apparent magnitudes $K_s=10.50$ and $K_s=11.07$) with high confidence. At $d=3.0$~kpc, the control field tells us that number of ``missing" giant branch stars in the selected region is $5.8 \pm 2.4$ and the Poisson probability of a gap at $10.50<K_s<11.07$ is $P(0)=0.3$\%. We adopt a likely distance range of 2.3 to 3.0~kpc to the dark cloud, with a preferred value of 2.6~kpc. This is consistent with kinematic distances slightly in excess of 3~kpc because peculiar velocities of only a few km/s relative to the Galactic rotation curve would modify the kinematic distance by several hundred pc. Inspection of the Gaia-based distances of \citet{bailer-jones18} confirms that the distance to the dark cloud is at least 2.3~kpc. The most distant stars having parallax measurements above 5$\sigma$
significance have a lower distance bound of c. 2.5 to 2.6~kpc in that work, at 90\% confidence. If we allow for the known negative offset of order $30$~$\mu$as in the current Gaia parallaxes \citep{lindegren18}, this translates to a minimum distance of about 2.3~kpc.

\section{Discussion}

The $\sim$8~mag mid-infrared amplitude of the outburst is higher than that of any previously recorded episodic accretion event, in any waveband.
This extreme value, together with the $\sim$1 decade duration and the mid-infrared location of the peak in the SED provide some potentially useful constraints
regarding which disc instability mechanism is operating here. In many previously known events, infrared amplitudes were slightly lower than optical
amplitudes but an amplitude in $K$ approaching 6~mag was seen in the embedded V346~Nor system in 2010-11 \citep{kospal20, kraus16}, during a 
brief dip between outbursts. Most other events in embedded YSOs have amplitudes of 2 to 4~mag in $K$ \citep[see][]{reipurth04, persi07, caratti11, connelley18},
though the eruptions of the class I YSOs OO~Ser \citep{hodapp96} and UKIDSS~J185318.36+012454.5 (hereafter UGPS~1853+0124, \citealt*{nikoghosyan17}) had amplitudes 
$\Delta$K=4.6~mag and $\Delta$K$\ge$4.8~mag, respectively. With a total duration of over a decade \citep{kospal07}, a very red 1--5~$\mu$m SED and a featureless, steeply 
rising 2~$\mu$m spectrum, the OO Ser event is the closest previous analogue to WISE~1422-6115, despite some differences (see below). Previously published YSO outbursts 
discovered by the VVV survey had $K_s$ amplitudes below $\sim$4~mag \citep[e.g.][]{cp17a, cp17b, medina18, teixeira18} but events with $K_s$ amplitudes up to 6.7~mag have 
been discovered more recently (Lucas et al., in prep.). WISE~1422-6115 appears to be an outlier but not by so much that we need to consider it an entirely different type of event.

\subsection{Accretion rate}

The accretion rate during outburst can be estimated from the luminosity using the equation $L$~=~$GM\dot{M}/2R_{*}$ \citep{shakura73}. The
1.25--22~$\mu$m luminosity was $L$~=~178--418~$(d/2.6~\mathrm{kpc})^2$~L$_{\odot}$ in 2010 (section \ref{sec:burst_sed}).
The low pre-outburst luminosity indicates a low mass progenitor but the stellar radius and mass are not known. From inspection of a sample
of 22 classical T Tauri stars with ages $<$3~Myr in \citet{grankin18} we find a mean mass of 0.65~M$_{\odot}$, a mean radius of 2.04~R$_{\odot}$ and
mean $R/M=3.9$~R$_{\odot}/$M$_{\odot}$, with a standard deviation of 1.9~R$_{\odot}/$M$_{\odot}$. Adopting the mean value of $R/M$, we find 
$\dot{M}$~=~4.4--10.3~$\times$~$10^{-5}$~M$_{\odot}$~yr$^{-1}$.
Making some allowance for flux emitted at $\lambda > 22~\mu$m and the slightly brighter infrared 2.15--4.6~$\mu$m fluxes in 2011-14 would bring the middle
of this range to approximately $\dot{M}=10^{-4}$~M$_{\odot}$~yr$^{-1}$. A luminosity of order $10^2$~L$_{\odot}$ is typical of a classical FUor
\citep{connelley18} and the implied accretion rate is therefore also typical of the rates inferred in the literature for such events (1--30$\times 10^{-5}$ M$_{\odot}$~yr$^{-1}$, \citealt{bell95}).

\subsection{Models and veiling}

Models employing the thermal-viscous instability mechanism \citep{bell94, bell95, lin85, hartmann85} naturally produce a high temperature (6000~K to 10~000~K) 
outburst in the innermost part of the disc (sub-au scales), leading to a lower amplitude in the infrared than at optical wavelengths. The detailed models of \citet{bell95} 
were able to reproduce many observed features of optically bright classical FUors, including a lower infrared amplitude, but some thought is needed to 
reconcile them with the mid-infrared outburst discussed here. The event duration can be adjusted in these models from $\sim$100~yr to $\sim$10~yr by increasing 
the adopted viscosity parameter, as noted by \citet{kospal07} for the case of OO Ser. Another class of model is the magneto-rotational instability (MRI) models, 
where MRI is triggered by spiral density waves created by gravitational instability in the outer disc, e.g. \citet{bae14, kadam20}. Again, these models lead to a
temperature of several thousand K in the disc.\footnote{In MRI models with a magnetically layered disc the temperature might be lower \citep{kadam20} but a low MRI ignition 
temperature was pre-defined in these models, constraining the temperature of the outburst.}

The magnetic gating model \citet{dangelo10, dangelo12} involves an instability near the co-rotation radius of the disc. For a typical YSO 
rotation period of 3~days this is 0.04~au for a 1~M$_{\odot}$ star, far smaller than the radius of a few au derived in section \ref{sec:burst_sed}.
However, the scenario of fragmentation in the outer disc, followed by inward migration to the inner disc and infall on to the star 
\citep{vorobyov06, vorobyov10, vorobyov15} can more naturally explain a release of energy at a radius of a few au and a high temperature is not necessarily
required, though this is still under investigation. In \citet*[][see figure 7 of that work]{vorobyov13}, we see an example of a mid-infrared peak in the SED arising from emission by 
a clump that has been scattered into the inner disc, though this is not in itself a burst event.

In the thermal instability and MRI models, the innermost part of the disc should become heated to several thousand K
as the mass flux through that region increases by orders of magnitude above the quiescent rate. 
Emission by this hot matter should be observed, as is seen in classical FUors, unless it is entirely obscured from view. A possible explanation is that the disc plane 
lies close to the line of sight so that the observed mid-infrared emission arises from reprocessed radiation further out in the disc. In this case much of the outburst region (perhaps 
all of it) would be closer to the star than the 4.6~au radius implied by the observed luminosity and the $\sim$800~K temperature. We might expect the SED to rise
from 4.6--22~$\mu$m in such a case, as opposed to the fall that is observed, due to reprocessing further out in the disc \citep{johnstone13} but this would depend on the
details of the disc structure and any effects such as disc shadowing or disc gaps.
A scattered light peak at 4.6~$\mu$m should also be considered but it seems unlikely, in view of the low albedo of sub-micron-sized dust grains at that wavelength.

Another possibility, perhaps less likely, is that the optical depth of the upper layers of the disc in the outburst region is sufficient to entirely obscure the hotter mid-plane layers from view (despite thermal inflation of the disc scale height) causing a shallow temperature gradient. A generally low disc surface temperature and a shallow temperature gradient may also occur in a new class of models where the burst is caused by an increase in turbulent viscosity due to vertical convection and convective transfer of energy \citep{pavlyuchenkov20}. A shallow 
temperature gradient helps to explain the almost featureless 2~$\mu$m spectrum, though we note that it was taken 6~mag below peak 2~$\mu$m brightness so gradients are 
expected to be shallower, weakening absorption features in the manner observed by \citet{guo20}.

\begin{figure}
	\vspace{0cm}
	\includegraphics[width=0.47\textwidth]{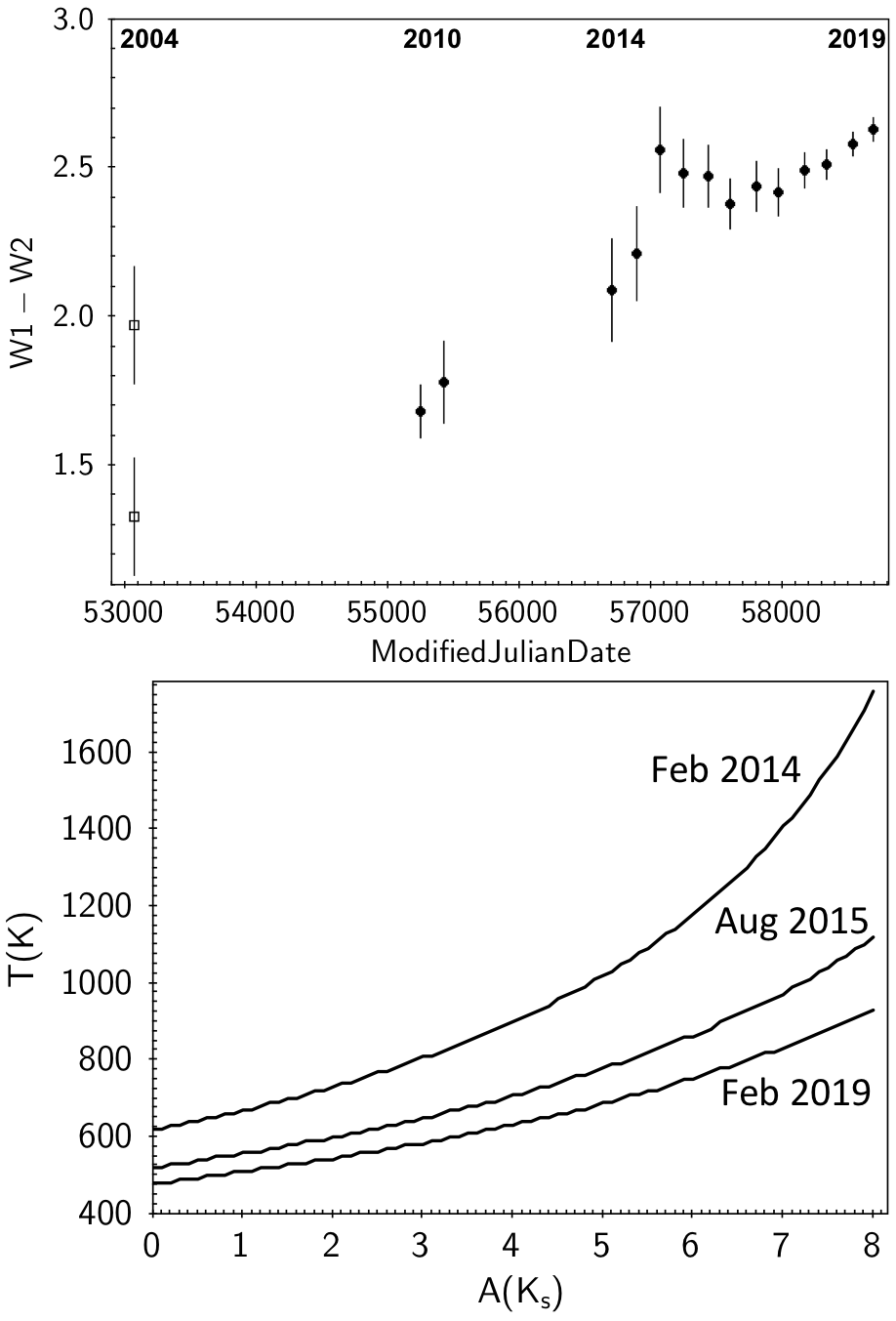}
	\vspace{0cm}
    \caption{Colour and temperature evolution. {\it (Upper panel):} the $W$1-$W$2 colour vs time. The GLIMPSE $I1$-$I2$ colour, measured pre-outburst, is plotted as two open squares:
    the lower square showing the published colour and the upper square derived after transformation to the WISE photometric system. During outburst, the overall trend was an increasingly 
    red colour as the outburst first brightened slightly from 2010--2014 and then faded thereafter. {\it (Lower panel):} best-fit temperature derived from the 2.15, 3.4 and 4.6~$\mu$m fluxes as 
    a function of assumed extinction, plotted at three separate epochs. A cooling trend is apparent as the outburst faded.}
    \label{fig:colevol}
\end{figure}

In comparison with most other outbursts in other embedded YSOs, WISE~1422-6115 event is redder and apparently cooler. The embedded YSO outburst events
V2775 Ori, V1647~Ori and GM~Cha have near infrared colours $J-K = 3.7$, 3.7 and 5.2 respectively, and their infrared spectra have temperatures similar to 
reddened M or L dwarf photospheres ($T$=3200~K was fitted for V2775~Ori by \citealt{caratti11}). The 800-1000~K temperature of WISE~1422-6115 in 2010 
(section \ref{sec:burst_sed}) is more T dwarf-like and it appears to have cooled towards the Y dwarf range since that time (see Figure \ref{fig:colevol}).
With $J-K_s=6.9, H-K_s=3.5$ in outburst and a 2~$\mu$m continuum spectrum that has a positive second derivative (see Figure \ref{fig:spec}), WISE~1422-6115 is amongst the
reddest events, alongside OO Ser, the massive YSO outburst of S255~IR~NIRS3 \citep{caratti17}, UGPS~1853+0124 and V723 Car \citep*{tapia15}.
None of the VVV YSO spectra in \citet{cp17b} have such steeply rising 2~$\mu$m continua as WISE~1422-6115 and OO Ser. Despite the similarities of the
two, OO Ser differs from WISE~1422-6115 in that the outburst of the former faded in a wavelength-independent manner and the outburst luminosity was relatively low at 
$\sim$30~L$_{\odot}$ \citep{kospal07}.

We note that UGPS~1853+0124 differed
from WISE~1422-6115 in that it has stayed at peak mid-infrared brightness for at least 10 years and showed (EXor-like) CO overtone emission bands.
The class I V723~Car event \citep{tapia15} had a more complex light curve and an EXor-like spectrum, with both Br$\gamma$ and CO overtone emission features.
\citet{tapia15} made a similar argument for an edge-on orientation to help explain the very red colour, $H-K=4.0$ at the peak, though very high foreground extinction also 
contributed. The V723~Car and UGPS~1853+0124 events may have been in relatively massive YSOs, though this is not entirely clear.\footnote{The distance, luminosity
and mass of UGPS~1853+0124 are uncertain \citep{nikoghosyan17}. A high mass, $M \sim$10~M$_{\odot}$, was claimed for V723~Car \citep{tapia15} but the mass 
and luminosity values given might be overestimated, having employed either mid-infrared data taken during the outburst or the far IR/submm flux of a 1700~M$_{\odot}$ 
``massive envelope" that may be due to a self-luminous cloud core rather than reprocessing.}

\subsection{Evolution of the event}
\label{sec:evol}

As noted earlier, the outburst declined more slowly with increasing wavelength, in the 1.6--4.6~$\mu$m range. In Figure \ref{fig:colevol} (upper panel) we plot the 
change in $W1$-$W2$ colour vs time. The GLIMPSE $I1$-$I2$ pre-outburst colour in 2004 (also plotted for comparison) is similar to the WISE colour
measured in 2010 if we allow for the difference between the {\it Spitzer/IRAC} and WISE bandpasses \citep{antoniucci14}. However, the outburst became 
redder between 2010 and the time of the brightest WISE measurements in 2014, then reddened further as the outburst began to fade, showing a blip in 2015 but
a clear trend to the red from late 2016 to late 2019. The ``redder when brighter" change between 2010 and 2014 suggests that the outburst was propagating outward
through the disc from 2010 to 2014, cooling slightly but increasing in luminosity as the area of the outbursting region increased. (We noted previously that the negative 
gradient of the SED from 4.6--22~$\mu$m in 2010 was a sign that energy had not yet been efficiently redistributed outward in the disc by that date).
This behaviour is seen in thermal instability models and in the MRI outburst models of \citet{kadam20}, though not in infalling fragment models. The data
are therefore perhaps most consistent with thermal instability or MRI models where the hot matter is veiled from view.

The behaviour as the outburst faded after 2014 can be understood as simple cooling of the outbursting region and perhaps further outward propagation.
In the lower panel of Figure \ref{fig:colevol} we plot the results of minumum $\chi^2$ fits to extinguished Planckian spectra, using the fluxes at 2.15, 3.4 and 4.6~$\mu$m. (We
interpolated the $K_s$ light curve using cubic spline fitting to estimate the fluxes at the WISE epochs.) For any extinction value in the range $0<A(K_s)<8$ there 
is a temperature that gives a minimum value of $\chi^2$. The best fit temperature rises slowly as a function of the adopted extinction (shown as the curves for three 
separate epochs) but the $\chi^2$ parameter varies by only $\sim25\%$ along each curve. Whilst the temperature at each epoch is only loosely constrained in these three point fits, 
a clear cooling trend is apparent. An isothermal fading due to increasing extinction alone appears unlikely since it would require very large changes in extinction.

\subsection{The cometary nebula}
Finally, the large size of the cometary reflection nebula requires some explanation. We have interpreted it as reflection from the massive cloud core 
(section \ref{sec:nebula}) which implies that the morphology must be due to the structure of the core. This may indicate that an outflow has cleared a low 
density cavity that extends through the core and out into the more diffuse interstellar medium. Alternatively, it may simply indicate a pre-existing density gradient 
in the cloud core that causes regions of lower optical depth to appear better illuminated. Narrow band  imaging in the 2.12~$\mu$m H$_2$
line or radio mapping in CO could distinguish between these two options. We have suggested a nearly edge-on disc orientation as an explanation for the
the mid-infrared flux peak, so if the structure is caused by an outflow we might have expected a bipolar cavity. However, high resolution imaging has shown that
the inner disc of a YSO can be highly inclined with respect to the outer disc \citep[see e.g.][]{benisty18} so the outflow cavity interpretation cannot be ruled out.

\section{Conclusions}

We have detected 23 new highly variable mid-infrared sources ($>$2~mag variation) in a catalogue of
WISE/NEOWISE sources projected in the vicinity of IRDCs. The majority of these relatively bright and red stars are 
YSOs but almost half are periodic, hence more likely to be dusty Mira-like variables. Since
we have searched only a very small portion of the Galactic plane, this demonstrates the potential of the 
ongoing NEOWISE mission for eruptive YSO science, complementing VVV/VVVX.

The most notable discovery is the $\sim$8~mag outburst in WISE~1422-6115, a low mass
YSO at a distance $d=2.3$--$3.0$~kpc that is presumed to have undergone an episodic accretion
event. In evolutionary terms, we classify it as a probable stage 1 protostar. 
The in-outburst luminosity of a few $\times10^2$~L$_{\odot}$ is higher than that of the surrounding 
massive cold cloud core so it is possible that the core was measurably warmed by the event, though
probably not enough to cause major changes to the cloud's chemical make-up. By contrast, the accretion disc 
and any surrounding circumstellar envelope would have been heated far more strongly.

Details of the event and the conclusions we have drawn are as follows.
\begin{enumerate}
\item GLIMPSE, MIPSGAL, WISE and VVV data indicate that the outburst began between 2004 and 2006
and the duration now exceeds 13~yr. After making some allowance for extinction and reprocessing, the luminosity is 
consistent with an accretion rate approaching $10^{-4}$~M$_{\odot}$yr$^{-1}$, comparable to a classical FUor event. 

\item The 4.6~$\mu$m peak in the SED corresponds to a temperature $T$~$\approx$~800--1000~K in 2010,
appearing to cool further as the outburst progressed and then faded. The 800-1000~K temperature and the
measured mid-infrared luminosity imply a disc radial location $R \approx 4.5$~au for the emitting region. The
apparent cooling as the outburst brightened slightly from 2010--2014 suggests an outward radial progression.
This contrasts with the YSO outburst of Gaia~17bpi \citep{hillenbrand18}, where there appears to
have been an inward progression. This disparate behaviour could be seen as evidence that different
disc instability mechanisms operated in the two outbursting systems.

\item The apparent lack of emission from hotter matter in the innermost disc is unusual, given that the innermost parts
of the disc are expected to be heated to several thousand K in either the thermal-viscous instability model 
\citep{bell94, bell95} or in MRI models \citep{bae14, kadam20}, with outward radial progression in both cases.
An F-type disc photosphere with supergiant gravity is observed at optical wavelengths in classical FUor events.
We suggest that the hot innermost regions of the disc may be obscured from view by a near edge-on disc orientation.
The 4.6~$\mu$m peak might then arise in large part from reprocessing of radiation emitted close to the star,
though there could also be a contribution from a radially extensive outburst region in the disc, with a spread of temperatures. 
Perhaps less likely, if the upper layers of the disc are denser and more opaque than in classical FUors, then the 
hot mid-plane layer where viscous energy is dissipated could be hidden. 

\item In the disc fragmentation/infalling fragments model a mid-infrared peak can arise naturally due to emission
from an infalling clump in the inner disc \citep{vorobyov13}. However, this is not in itself a burst event and radially outward 
progression of a burst would be unexpected. In the context of the disc fragmentation model, the low pre-outburst luminosity of the 
protostar and the location inside a cold dense cloud core should lead to a low temperature in the outer disc that would tend to 
facilitate fragmentation, as noted by \citet{vorobyov10}. 

\item The very high amplitude, more than 13~yr duration and mid-infrared location disfavour the magnetic gating mechanism, 
which was designed more to explain EXor-like events, \citep[see][]{dangelo10,dangelo12}.
\end{enumerate}

Further observations of this system are certainly desirable to better understand the event. Very high resolution
submm continuum imaging with ALMA might be able to detect signs of fragmentation in the outer disc. Mid-infrared
spectroscopy (see \citealt{kospal20b}) and interferometry should be practicable, in view of the slow rate of decline of the outburst 
in that waveband. Sub-arcsecond millimetre waveband data would be able to detect the expected circumstellar envelope within the 
cloud core, confirming stage I evolutionary status. Either $^{13}$CO mapping or 2.12~$\mu$m H$_2$ imaging 
could be used to search for an outflow, investigate the arc of nebulosity and better understand the nature of the 
large cometary reflection nebula.

Finally, another of the new discoveries, WISEA J185720.27+015711.8 (source 21 in Table \ref{tab:table1}) underwent
a mid-infrared outburst of 4.6~mag and it has remained in a bright state for several years. This unusually high amplitude event 
also warrants further investigation.

\section*{Acknowledgements}

We thank the referee for reading the paper and encouraging us to provide more detail in several places.
This publication makes use of data products from the WISE satellite, which is a joint project of the University of California, Los Angeles, and the Jet Propulsion Laboratory/California Institute of Technology, funded by the National Aeronautics and Space Administration (NASA). 
The work is based in part on observations obtained at the Southern Astrophysical Research (SOAR) telescope, which is a joint
project of the Minist\'{e}rio da Ci\^{e}ncia, Tecnologia e Inova\c{c}\~{o}es (MCTI) do Brasil, the US National
Science FoundationÕs National Optical-Infrared Astronomy Research Laboratory (NOIRLab), the University of North
Carolina at Chapel Hill (UNC), and Michigan State University (MSU).
This research has made use of the NASA/IPAC Infrared Science Archive, which is funded by the National Aeronautics and Space Administration and operated by the California 
Institute of Technology.
We also acknowledge use of NASA's Astrophysics Data System Bibliographic Services and the SIMBAD database operated at CDS, Strasbourg, France. 
We also made use of the VizieR catalogue access tool, CDS, Strasbourg, France (DOI : 10.26093/cds/vizier). The original description of the VizieR service was published in 
2000, A\&AS 143, 23. PWL and ZG acknowledge support by STFC Consolidated Grant ST/R00905/1 and CM is supported by an STFC studentship funded by grant ST/S505419/1.
We gratefully acknowledge data from the ESO Public Survey program ID 179.B-2002 taken with the VISTA telescope, and products from the Cambridge Astronomical Survey Unit (CASU). EV acknowledges support from the Russian Fund for Fundamental Research, Russian-Taiwanese project 19-52-52011.
DM acknowledges support from the FONDECYT Regular grant No. 1170121,  the BASAL Center for Astrophysics and Associated Technologies (CATA) through grant AFB170002.
RKS acknowledges support from CNPq/Brazil through project 305902/2019-9.\\

\noindent{\bf Data Availability}\\
The WISE, {\it Spitzer} and 2MASS data underlying this article are publicly available at \url{https://irsa.ipac.caltech.edu/Missions/wise.html}, \url{https://irsa.ipac.caltech.edu/Missions/spitzer.html} and \url{https://irsa.ipac.caltech.edu/Missions/2mass.html} respectively. The VVV and VVVX data are publicly available at the ESO archive 
\url{http://archive.eso.org/cms.html}. The relevant reduced products for the most recent VVVX epochs have not yet been publicly released but are available on request to the first 
author. {\it Akari} photometry are available at \url{https://vizier.u-strasbg.fr/viz-bin/VizieR-2}. {\it Herschel} data are available at the Herschel Science Archive, 
\url{http://archives.esac.esa.int/hsa/whsa/}. UKIDSS data are available at the WFCAM Science Archive \url{wsa.roe.ac.uk}.
The SOAR data are available at \url{http://star.herts.ac.uk/~pwl/Lucas/SOAR/}.

\bibliographystyle{mnras}
\bibliography{wv2bib} 



\appendix
\section{Additional Photometry}
\label{sec:A}

The VVV $K_s$ light curve consists of 157 data points, after discarding images with seeing $>$1.2\arcsec and a few highly discrepant data points. 
There were four contemporaneous image stacks at each epoch (from adjoining VVV tiles d051 and d052) each showing numerous adjacent stars 
for visual comparison, which made it easy to reject the outlying data. 

A difficulty with the VVV/VVVX data is that the source was not only saturated  in $K_s$ at maximum brightness (i.e. in 2010-2013 and in a minority of 
images from 2014) but also spatially extended on a scale of a few arc seconds, though the flux profile is strongly peaked, see 
Figure \ref{fig:VVVims}. 
The saturation level varies between the 16 VISTA/VIRCAM arrays. 
Inspection of the radial intensity profile of the source in VISTA pawprint images from the 4 arrays with detections showed
that the profile just outside the saturated core resembles that of a point source. From the photometric apertures provided by the
Cambridge Astronomical Survey Unit (CASU) VISTA pipeline we used apertures 3 and 4 (diameters of 2\arcsec~and 2$\sqrt{2}$\arcsec, respectively) 
to compute a saturation-corrected magnitude in a similar manner to that described by \citet{irwin09}.This was done for the 2010-2013
data, with an estimated uncertainty of 0.2~mag, and visual comparison with saturated stars having 2MASS photometry was used to
check that the results were sensible. Saturated data from 2014 were discarded because the image profile had changed.
For unsaturated $K_s$ data from 2014 onward, and for the other VVV passbands, we give point source photometry in aperture 3. 
This will slightly underestimate the total brightness of the source but it provides a reasonably consistent measurement of the brightness of 
central flux peak, for comparison with the earlier data.

The integrated VVV magnitudes of the saturated $K_s$ March 2010 photometry used for SED fitting (see section \ref{sec:SED}) were calculated by
computing the difference between the fluxes measured by the CASU pipeline in apertures 7 (8\arcsec~diameter) and aperture 3 and adding this quantity
to the saturation-corrected aperture 3 flux computed earlier. After converting the resulting flux value to a magnitude, we added the aperture correction 
for aperture 3 and subtracted the aperture correction for aperture 7 to arrive at an aperture corrected magnitude in aperture 7. The aperture corrections
were provided by the CASU pipeline for each image.

\begin{table}
	\begin{centering}
	\caption{Additional time series photometry for WISE~1422-6115}
	\label{tab:A1}
	\begin{tabular}{lcccc} \hline
Dataset & MJD & Bandpass & Magnitude & Error\\
DENIS & 51014.69 & $K_s$ & $>$13.6 & \\
2MASS & 51615.29 & $K_s$ & $>$15.0 & \\
GLIMPSE & 53075.39 & [3.6] & 14.41 & 0.13 \\
GLIMPSE & 53075.56 & [4.5] & 13.08 & 0.15 \\
GLIMPSE & 53075.39 & [5.8] & 12.44 & 0.17 \\
GLIMPSE & 53075.56 & [8.0] & 11.57 & 0.32 \\
MIPSGAL & 53834.33 & [24] & 4.74 & 0.03 \\
{\it Akari}/IRC & -  &  [9] & 5.90 & 0.35 \\
{\it Akari}/IRC & -  &  [18] & 3.83 & 0.03 \\
VVV/VVVX & 55274.245 & $Z$ & $>$20.0 & \\
VVV/VVVX & 55281.254 & $Y$ & 19.14 & 0.10 \\
VVV/VVVX & 55281.255 & $Y$ & 19.42 & 0.15 \\
VVV/VVVX & 55282.162 & $Y$ & 19.44 & 0.16 \\
VVV/VVVX & 55282.163 & $Y$ & 19.55 & 0.21 \\
VVV/VVVX & 55274.253 & $J$ & 16.07 & 0.02 \\
VVV/VVVX & 55274.255 & $J$ & 16.06 & 0.02 \\
VVV/VVVX & 55288.293 & $J$ & 16.11 & 0.02 \\
VVV/VVVX & 55288.295 & $J$ & 16.11 & 0.02 \\
VVV/VVVX & 55274.245 & $H$ & 12.65 & 0.02 \\
VVV/VVVX & 55288.285 & $H$ & 12.68 & 0.02 \\
VVV/VVVX & 57176.178 & $H$ & 16.30 & 0.03 \\
VVV/VVVX & 57176.179 & $H$ & 16.27 & 0.04 \\
VVV/VVVX & 57181.250 & $H$ & 16.34 & 0.05 \\
VVV/VVVX & 57181.252 & $H$ & 16.26 & 0.05 \\
VVV/VVVX & 55260.34247 & $K_s$ & 9.08 & 0.20 \\
VVV/VVVX & 55260.3431 & $K_s$ & 9.19 & 0.20 \\
VVV/VVVX & 55260.34724 & $K_s$ & 9.27 & 0.20 \\
VVV/VVVX & 55260.34796 & $K_s$ & 9.16 & 0.20 \\
VVV/VVVX & 55262.39191 & $K_s$ & 9.20 & 0.20 \\
VVV/VVVX & 55262.39336 & $K_s$ & 9.25 & 0.20 \\
VVV/VVVX & 55262.39404 & $K_s$ & 9.16 & 0.20 \\
VVV/VVVX & 55264.39126 & $K_s$ & 9.21 & 0.20 \\
VVV/VVVX & 55264.392 & $K_s$ & 9.26 & 0.20 \\
VVV/VVVX & 55264.39267 & $K_s$ & 9.23 & 0.20 \\
VVV/VVVX & 55274.24962 & $K_s$ & 9.18 & 0.20 \\
VVV/VVVX & 55274.36398 & $K_s$ & 9.05 & 0.20 \\
VVV/VVVX & 55274.3646 & $K_s$ & 9.16 & 0.20 \\
VVV/VVVX & 55274.36793 & $K_s$ & 9.18 & 0.20 \\
VVV/VVVX & 55274.36857 & $K_s$ & 9.17 & 0.20 \\
VVV/VVVX & 55283.24924 & $K_s$ & 9.08 & 0.20 \\
VVV/VVVX & 55283.25274 & $K_s$ & 9.23 & 0.20 \\
VVV/VVVX & 55283.25337 & $K_s$ & 9.10 & 0.20 \\
VVV/VVVX & 55288.28806 & $K_s$ & 9.22 & 0.20 \\
VVV/VVVX & 55288.28954 & $K_s$ & 9.07 & 0.20 \\
VVV/VVVX & 55778.06854 & $K_s$ & 8.52 & 0.20 \\
VVV/VVVX & 55778.07018 & $K_s$ & 8.62 & 0.20 \\
VVV/VVVX & 55778.07102 & $K_s$ & 8.54 & 0.20 \\
VVV/VVVX & 55780.07017 & $K_s$ & 8.68 & 0.20 \\
VVV/VVVX & 55785.99496 & $K_s$ & 8.72 & 0.20 \\
VVV/VVVX & 55785.99667 & $K_s$ & 8.76 & 0.20 \\
VVV/VVVX & 55795.02149 & $K_s$ & 8.56 & 0.20 \\
VVV/VVVX & 55795.02306 & $K_s$ & 8.72 & 0.20 \\
VVV/VVVX & 55803.98976 & $K_s$ & 8.34 & 0.20 \\
VVV/VVVX & 55803.99062 & $K_s$ & 8.39 & 0.20 \\
VVV/VVVX & 55803.99142 & $K_s$ & 8.53 & 0.20 \\
VVV/VVVX & 55808.99112 & $K_s$ & 8.27 & 0.20 \\
VVV/VVVX & 55808.99192 & $K_s$ & 8.27 & 0.20 \\
VVV/VVVX & 55808.99272 & $K_s$ & 8.35 & 0.20 \\
VVV/VVVX & 55995.18952 & $K_s$ & 8.64 & 0.20 \\
VVV/VVVX & 55995.19099 & $K_s$ & 8.73 & 0.20 \\
VVV/VVVX & 55995.19174 & $K_s$ & 8.69 & 0.20 \\
VVV/VVVX & 56014.27852 & $K_s$ & 8.59 & 0.20 \\
VVV/VVVX & 56014.28026 & $K_s$ & 8.96 & 0.20 \\
VVV/VVVX & 56014.28119 & $K_s$ & 8.60 & 0.20 \\
\end{tabular}
	\end{centering}\\
\end{table}

\begin{table}
	\begin{centering}
	\contcaption{}
	\begin{tabular}{lcccc} \hline
Dataset & MJD & Bandpass & Magnitude & Error\\
VVV/VVVX & 56058.30581 & $K_s$ & 8.72 & 0.20 \\
VVV/VVVX & 56058.30765 & $K_s$ & 8.82 & 0.20 \\
VVV/VVVX & 56371.21412 & $K_s$ & 9.02 & 0.20 \\
VVV/VVVX & 56371.21564 & $K_s$ & 9.10 & 0.20 \\
VVV/VVVX & 56371.21643 & $K_s$ & 8.85 & 0.20 \\
VVV/VVVX & 56372.20321 & $K_s$ & 8.88 & 0.20 \\
VVV/VVVX & 56372.20405 & $K_s$ & 9.01 & 0.20 \\
VVV/VVVX & 56372.20483 & $K_s$ & 8.93 & 0.20 \\
VVV/VVVX & 56372.20566 & $K_s$ & 8.91 & 0.20 \\
VVV/VVVX & 56373.19458 & $K_s$ & 8.99 & 0.20 \\
VVV/VVVX & 56373.19615 & $K_s$ & 9.04 & 0.20 \\
VVV/VVVX & 56373.19759 & $K_s$ & 8.89 & 0.20 \\
VVV/VVVX & 56374.19197 & $K_s$ & 9.06 & 0.20 \\
VVV/VVVX & 56375.25579 & $K_s$ & 8.59 & 0.20 \\
VVV/VVVX & 56375.2566 & $K_s$ & 8.61 & 0.20 \\
VVV/VVVX & 56375.25738 & $K_s$ & 8.65 & 0.20 \\
VVV/VVVX & 56375.25819 & $K_s$ & 8.47 & 0.20 \\
VVV/VVVX & 56419.25141 & $K_s$ & 8.57 & 0.20 \\
VVV/VVVX & 56419.25216 & $K_s$ & 8.70 & 0.20 \\
VVV/VVVX & 56419.25293 & $K_s$ & 8.76 & 0.20 \\
VVV/VVVX & 56428.3069 & $K_s$ & 9.06 & 0.20 \\
VVV/VVVX & 56448.22414 & $K_s$ & 9.08 & 0.20 \\
VVV/VVVX & 56448.22497 & $K_s$ & 9.14 & 0.20 \\
VVV/VVVX & 56448.22655 & $K_s$ & 9.10 & 0.20 \\
VVV/VVVX & 56453.21702 & $K_s$ & 9.02 & 0.20 \\
VVV/VVVX & 56453.21779 & $K_s$ & 8.85 & 0.20 \\
VVV/VVVX & 56453.21861 & $K_s$ & 8.92 & 0.20 \\
VVV/VVVX & 56453.2194 & $K_s$ & 8.78 & 0.20 \\
VVV/VVVX & 56460.99369 & $K_s$ & 8.91 & 0.20 \\
VVV/VVVX & 56460.99446 & $K_s$ & 8.99 & 0.20 \\
VVV/VVVX & 56460.99521 & $K_s$ & 9.12 & 0.20 \\
VVV/VVVX & 56460.99594 & $K_s$ & 8.91 & 0.20 \\
VVV/VVVX & 56463.17171 & $K_s$ & 9.21 & 0.20 \\
VVV/VVVX & 56463.17253 & $K_s$ & 9.24 & 0.20 \\
VVV/VVVX & 56463.17329 & $K_s$ & 9.15 & 0.20 \\
VVV/VVVX & 56463.17412 & $K_s$ & 9.10 & 0.20 \\
VVV/VVVX & 56466.18514 & $K_s$ & 9.16 & 0.20 \\
VVV/VVVX & 56466.18668 & $K_s$ & 9.23 & 0.20 \\
VVV/VVVX & 56466.18747 & $K_s$ & 9.14 & 0.20 \\
VVV/VVVX & 56467.16656 & $K_s$ & 8.96 & 0.20 \\
VVV/VVVX & 56467.1674 & $K_s$ & 9.05 & 0.20 \\
VVV/VVVX & 56467.16816 & $K_s$ & 9.09 & 0.20 \\
VVV/VVVX & 56468.16904 & $K_s$ & 9.07 & 0.20 \\
VVV/VVVX & 56468.16983 & $K_s$ & 9.18 & 0.20 \\
VVV/VVVX & 56468.17057 & $K_s$ & 9.17 & 0.20 \\
VVV/VVVX & 56468.17137 & $K_s$ & 9.05 & 0.20 \\
VVV/VVVX & 56469.12137 & $K_s$ & 9.10 & 0.20 \\
VVV/VVVX & 56813.04202 & $K_s$ & 11.03 & 0.02 \\
VVV/VVVX & 56813.04381 & $K_s$ & 10.94 & 0.02 \\
VVV/VVVX & 56813.05533 & $K_s$ & 11.00 & 0.02 \\
VVV/VVVX & 56813.0571 & $K_s$ & 10.92 & 0.02 \\
VVV/VVVX & 56813.96544 & $K_s$ & 10.94 & 0.02 \\
VVV/VVVX & 56813.96625 & $K_s$ & 10.96 & 0.02 \\
VVV/VVVX & 56813.96822 & $K_s$ & 10.96 & 0.02 \\
VVV/VVVX & 56813.97878 & $K_s$ & 10.94 & 0.02 \\
VVV/VVVX & 56813.97961 & $K_s$ & 11.02 & 0.02 \\
VVV/VVVX & 56813.98135 & $K_s$ & 10.95 & 0.02 \\
VVV/VVVX & 56813.9912 & $K_s$ & 10.95 & 0.02 \\
VVV/VVVX & 56813.99204 & $K_s$ & 11.01 & 0.02 \\
VVV/VVVX & 56813.99388 & $K_s$ & 10.96 & 0.02 \\
VVV/VVVX & 56814.00341 & $K_s$ & 10.94 & 0.02 \\
VVV/VVVX & 56814.00422 & $K_s$ & 11.00 & 0.02 \\
VVV/VVVX & 56814.00643 & $K_s$ & 10.98 & 0.02 \\
VVV/VVVX & 56814.04191 & $K_s$ & 10.93 & 0.02 \\
\end{tabular}
	\end{centering}\\
\end{table}

\begin{table}
	\begin{centering}
	\contcaption{}
	\begin{tabular}{lcccc} \hline
VVV/VVVX & 56814.04274 & $K_s$ & 10.95 & 0.02 \\
VVV/VVVX & 56814.04373 & $K_s$ & 10.95 & 0.02 \\
VVV/VVVX & 56814.04456 & $K_s$ & 10.96 & 0.02 \\
VVV/VVVX & 56825.01277 & $K_s$ & 11.00 & 0.02 \\
VVV/VVVX & 56825.01364 & $K_s$ & 11.03 & 0.02 \\
VVV/VVVX & 57118.15631 & $K_s$ & 12.03 & 0.02 \\
VVV/VVVX & 57118.15709 & $K_s$ & 12.05 & 0.02 \\
VVV/VVVX & 57134.0828 & $K_s$ & 12.10 & 0.02 \\
VVV/VVVX & 57150.03191 & $K_s$ & 12.08 & 0.02 \\
VVV/VVVX & 57150.03299 & $K_s$ & 12.08 & 0.02 \\
VVV/VVVX & 57150.0339 & $K_s$ & 12.11 & 0.02 \\
VVV/VVVX & 57150.03507 & $K_s$ & 12.10 & 0.02 \\
VVV/VVVX & 57165.0114 & $K_s$ & 12.09 & 0.02 \\
VVV/VVVX & 57165.01219 & $K_s$ & 12.16 & 0.02 \\
VVV/VVVX & 57165.01362 & $K_s$ & 12.14 & 0.02 \\
VVV/VVVX & 57165.01452 & $K_s$ & 12.14 & 0.02 \\
VVV/VVVX & 57171.02674 & $K_s$ & 12.10 & 0.02 \\
VVV/VVVX & 57171.02757 & $K_s$ & 12.16 & 0.02 \\
VVV/VVVX & 57171.02829 & $K_s$ & 12.14 & 0.02 \\
VVV/VVVX & 57171.02911 & $K_s$ & 12.14 & 0.02 \\
VVV/VVVX & 57172.05397 & $K_s$ & 12.10 & 0.02 \\
VVV/VVVX & 57172.05474 & $K_s$ & 12.12 & 0.02 \\
VVV/VVVX & 57172.05543 & $K_s$ & 12.18 & 0.02 \\
VVV/VVVX & 57172.05623 & $K_s$ & 12.17 & 0.02 \\
VVV/VVVX & 57176.18236 & $K_s$ & 12.14 & 0.02 \\
VVV/VVVX & 57176.18403 & $K_s$ & 12.17 & 0.02 \\
VVV/VVVX & 57181.25722 & $K_s$ & 12.12 & 0.02 \\
VVV/VVVX & 57585.07992 & $K_s$ & 12.65 & 0.02 \\
VVV/VVVX & 57585.0807 & $K_s$ & 12.68 & 0.02 \\
VVV/VVVX & 57585.08142 & $K_s$ & 12.69 & 0.02 \\
VVV/VVVX & 57585.08222 & $K_s$ & 12.70 & 0.02 \\
VVV/VVVX & 57928.18046 & $K_s$ & 13.21 & 0.02 \\
VVV/VVVX & 57933.0333 & $K_s$ & 13.19 & 0.02 \\
VVV/VVVX & 57933.03393 & $K_s$ & 13.18 & 0.02 \\
VVV/VVVX & 57933.03461 & $K_s$ & 13.22 & 0.02 \\
VVV/VVVX & 57933.03533 & $K_s$ & 13.21 & 0.02 \\
VVV/VVVX & 58208.31044 & $K_s$ & 13.66 & 0.02 \\
VVV/VVVX & 58208.31118 & $K_s$ & 13.67 & 0.02 \\
VVV/VVVX & 58212.39832 & $K_s$ & 13.70 & 0.03 \\
VVV/VVVX & 58262.99078 & $K_s$ & 13.85 & 0.02 \\
VVV/VVVX & 58262.99156 & $K_s$ & 13.81 & 0.02 \\
VVV/VVVX & 58654.20043 & $K_s$ & 14.28 & 0.03 \\
VVV/VVVX & 58654.20121 & $K_s$ & 14.27 & 0.03 \\
VVV/VVVX & 58654.20198 & $K_s$ & 14.33 & 0.03 \\
VVV/VVVX & 58654.20271 & $K_s$ & 14.31 & 0.03 \\
VVV/VVVX & 58701.08159 & $K_s$ & 14.34 & 0.02 \\
VVV/VVVX & 58701.08231 & $K_s$ & 14.34 & 0.02 \\
VVV/VVVX & 58701.08303 & $K_s$ & 14.37 & 0.02 \\
VVV/VVVX & 58701.08378 & $K_s$ & 14.37 & 0.03 \\
VVV/VVVX & 58702.06429 & $K_s$ & 14.32 & 0.03 \\
VVV/VVVX & 58702.06503 & $K_s$ & 14.36 & 0.02 \\
VVV/VVVX & 58702.06659 & $K_s$ & 14.40 & 0.02 \\
VVV/VVVX & 58702.06742 & $K_s$ & 14.37 & 0.03 \\
\end{tabular}
	\end{centering}\\
\end{table}

\begin{table}
	\begin{centering}
	\caption{{\it Herschel}/Hi-Gal and ATLASGAL scaled photometry for HIGALBM~313.6699-0.3092}
	\label{tab:A2}
	\begin{tabular}{ccc} \hline
Wavelength ($\mu$m) & Flux (Jy) & Error (Jy)\\
70 & 3.12 & 0.14 \\
160 & 10.89 & 0.16 \\
250 & 14.66 & 0.41 \\
350 &  4.48 & 0.22 \\
500 &  2.37 & 0.15 \\
870 & 0.97 & 0.19 \\
\end{tabular}
	\end{centering}\\
\end{table}


\bsp	
\label{lastpage}
\end{document}